# How do ASA Ethical Guidelines Support U.S. Guidelines for Official Statistics?

Jennifer Park[1]
Rochelle E. Tractenberg[2]

1. Committee on National Statistics, National Academies of Science, Engineering, and Medicine, Washington, DC USA

2. Collaborative for Research on Outcomes and –Metrics; and Departments of Neurology; Biostatistics, Bioinformatics & Biomathematics; and Rehabilitation Medicine, Georgetown University, Washington, DC, USA

ORCID: 0000-0002-1121-2119

**Correspondence to:**

Rochelle E. Tractenberg
Georgetown University Neurology
Room 207, Building D
4000 Reservoir Rd., NW
Washington, DC, 20057 USA
**Email:** rochelle -dot- tractenberg -at -gmail -dot- com

Acknowledgement: There are no actual or potential conflicts of interest.

Running Head: ASA Ethical Guidelines support US Official Statistics

How do ASA Ethical Guidelines Support U.S. Guidelines for Official Statistics?

**Boxes**
1 Data sources examined by intended purpose and intended audience
2 ASA Ethical Guidelines for Statistical Practice, 2022
3 OMB Statistical Policy Directive 1, 2014
4 National Academies of Science Principles and Practices, 2021
5 OMB Data Ethics Tenets, 2020

A. Statement of the Issue

The past several years have been a time of great energy and enthusiasm for official statistics and data science.[2][3][4][5][6] Perhaps the most impactful of all federal laws affecting statistics and data for the public good, the Foundations for Evidence-Based Policymaking Act of 2018 (Evidence Act)[7] acknowledges the broad and growing community of statistical and data science practitioners. The implementation of the Evidence Act will occur through federal regulation and agency policy, establishing new relationships and expectations for practitioners of statistics and data science in federal agencies. As is customary, the public will be called to comment on any proposed new regulations.

Setting statistical and data science policy will take careful thought. For example, the Evidence Act has enacted "presumed access" of federal statistical agencies to other agencies' data for the purposes of statistical analysis. This authority changes the balance of trust in favor of relevance and timeliness because such access was not described to data providers at the time of initial collection. [8] The rapid release of statistics and data to inform COVID-19 response has also challenged priorities and practice regarding the balance of timeliness and accuracy. The CHIPS Act has enabled the development of a National Secure Data Service to improve public access to a wide range of data (particularly, blended data) for statistical purposes, which may challenge priorities and practices regarding the balance of accuracy in favor of timeliness inasmuch as blended data often leverage imperfect data linkages and inconsistent data values. Given these responses to new and urgent data needs, and the likelihood of other similar situations in the future, professional ethics are a natural resource to leverage in order to guide new - and reinforce existing - standards.

In 2022, the American Statistical Association revised its Ethical Guidelines for Statistical Practice. Originally issued in 1982, these Guidelines describe responsibilities of "the ethical statistical practitioner" to their profession, to their research subjects, as well as to their community of practice. These guidelines are intended as a framework to assist decision-making by statisticians working across academic, research, and government environments. For the first time, the 2022 Guidelines describe the ethical obligations of organizations and institutions that use statistical practice.

---

[2] See OMB M-13-13 memo, which calls for federal agency efforts to support transparency, accessibility and discoverability of data anticipating secondary uses. This also signals a shifting prioritization of privacy to accessibility.

[3] See OSTP 2013 memo directing agencies with more than 100 million in R and D funds to develop a plan to improve public access to federally funded research with a special focus on scholarly articles and digital data underpinning that research. The 2013 represented a shift toward data sharing and was interpreted as improving equity.

[4] See Commission on Evidence-Based Policymaking, The Promise of Evidence-Based-Policymaking (2017), available at https"://www.cep.gov/cep-final-report.html even more strongly urged secondary data access by "default" unless otherwise prohibited, to improve utility of federal data for policy making. A subsequent NASEM report described methods to balance accuracy and privacy in blended data; see Innovations in Federal Statistics: Combining Data Sources While Protecting Privacy (2017), available at https://www.nap.edu/catalog/24893/federal-statistics-multiple-data-sources-and-privacy-protecti6n-next-steps. Forthcoming consensus reports from NASEM explore emerging needs and practices for blended data to support US data infrastructure. Both bodies emphasize the importance of protecting personal information while improving access.

[5] See OMB M-19-15 memo to update the implementation of the information quality act regarding confidentiality and access. This document recognizes shifting priority to secondary uses of data (after initial consent), increasing accessibility, but continuing to guard privacy.

[6] See OSTP, August 25, 2022 re public access to federally-funded research and data without embargo and free.

[7] Pub. L. No. 115-435.

[8] Note that information collection conducted by several federal agencies for non-statistical purposes, such as IRS and SSA, are not required to obtain consent to share and link data for additional uses. Additionally, the Census Bureau is permitted access to administrative records for statistical purposes only, but does not require consent.

Of course, the U.S. has long-established guidance for federal statistical agencies. Prévost (2018) observes general agreement across sources for norms in statistical activities, particularly as they are reflected in –and have influenced--official statistics. These norms have emphasized "relevance, accuracy, timeliness, coherence and comparability, interpretability and accessibility" (Prévost, 2018) and are currently embodied in US federal guidance (see above)) as well as international guidance (e.g., UN Fundamental Principles of Official Statistics, OECD Good Statistical Practice, and European Statistics Code of Practice).

As will be discussed later in this paper, the ASA Ethical Guidelines and guidelines for US official statistics and data were developed with different objectives and therefore for different audiences. There are many purposeful differences among them. Yet, examining areas of alignment and difference may inform future applications and potential growth in both professional and official guidance. The more that statistical agencies embody such dimensions as relevance in their programs, the more likely that data users and stakeholders will have confidence in the quality of the statistics and in the statistical profession itself.

How do the ASA Ethical Guidelines support U.S. guidelines for official statistics and data? Are they largely complementary in scope and specificity, or do the ethical practice standards of the ASA omit or add new areas of guidance? Can understanding alignment among these source documents improve the U.S. practice of official statistics? Further, how would understanding this alignment inform federal statistical policy and anticipated implementation rules for the Evidence Act?

This paper examines alignment between the ASA Ethical Guidelines and other long-established normative guidelines for US official statistics: the OMB Statistical Policy Directives 1, 2, and 2a[9] NASEM Principles and Practices, and the OMB Data Ethics Tenets. Our analyses ask how the recently updated ASA Ethical Guidelines can support these guidelines for federal statistics and data science. The analysis uses a form of qualitative content analysis, the alignment model, to identify patterns of alignment, and potential for tensions, within and across guidelines. The paper concludes with recommendations to policy makers when using ethical guidance to establish parameters for policy change and the administrative and technical controls that necessarily follow.

1. Exploration of guidelines can support professional quality

An examination of how and where ASA Ethical Guidelines align with federal guidelines can identify shared practices and behaviors that are likely to improve the quality of statistics and statistical work. Conversely, there may be areas of misalignment that suggest ways to modify or clarify the ASA Ethical Guidelines or the official statistics guidelines, or both.

To begin this examination, we first address two potential sources of confusion that may arise when examining guidelines: the nesting of constructs and the selection of similar terms. For the purposes of this paper, Figure 1 describes a relationship among constructs within a given set of ethical guidelines. For example, ethical guidance may articulate goals, activities that achieve those goals, and the manner in which activities should be conducted. Figure 1 also notes where similar terms for these constructs are found among the guidelines examined in this paper.

---

[9] For reasons that will be discussed later in this paper, we focused our examination on Statistical Policy Directive 1, 2, and 2a, although Statistical Policy Directives 3 and 4 also provide important technical guidance to federal statistical agencies.

Guidance documents are often intended for different audiences and to serve different purposes. US federal statistics guidelines relate primarily to the goals of federal statistical work, and, for narrower audiences, on particular statistical practice. Technical guidance for particular topics, such as the conduct of statistical surveys, the release of federal principal economic indicators, and the release of other statistical products, for example, provide concrete direction on expected practices for certain federal statistical agencies conducting certain statistical products. However, these guidelines tend to remain silent on specifying how (sometimes competing) priorities should or could be balanced to provide the greatest benefit and least harm–i.e., ethical statistical conduct.

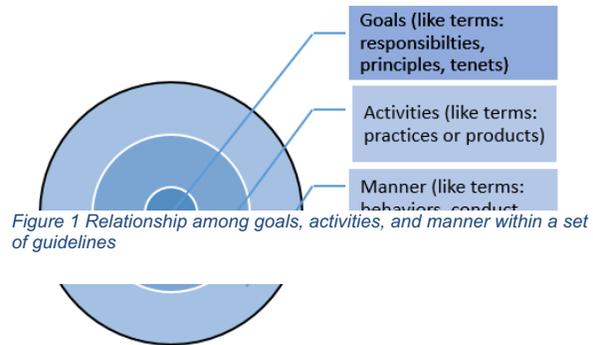

*Figure 1 Relationship among goals, activities, and manner within a set of guidelines*

Yet, the "quality" of statistical output is intimately tied to the expertise and ethical conduct of the practitioner, and may be responsive to contextual motivations and norms (Prévost, 2018). Müller et al. (2021, p. 40) and especially Tractenberg (2022-A, B) describe how these outputs are influenced by the orientation of the statistical practitioner to ethical principles and guidelines.[10] A code of ethics, Müller et al. argue, "includes principles applicable to all the ethical aspects of a professional's work, including its relation to the public interest" (p. 40).

Although federal statistical guidance was intended generally for a broad audience, its community has grown and diversified dramatically within the past twenty years. Currently, there are 13 principal federal statistical agencies and 3 recognized federal statistical units whose mission is to provide official federal statistics. However, there are over a hundred other statistical programs throughout the federal government that, as part of their mission, also collect data and produce statistics to inform federal policy. (Statistical Programs of the United States Government, 2019/2020) In addition, the 2018 Evidence Act defined roles for Chief Data Officers and Chief Evaluation Officers across 24 federal agencies party to the Chief Financial Officer Act. The contributions of this wider community of federal agencies to statistical practice was recognized by enlarging membership of the Interagency Council on Statistical Policy (ICSP), chaired by the US Chief Statistician.[11]

It may be the case that this community of professionals conducting statistical work have somewhat different objectives, and therefore, make decisions reasonably according to somewhat different professional frameworks. At the level of goals (for example, objectivity), there may be greater agreement and therefore less need for specificity in guidelines. It may also be the case that the manner of statistical work (for example, particular data product release procedures) may be too detailed for embedding in broad federal guidance. Nonetheless, it is at this detail level that statistics and data science professionals make decisions (for example, by balancing relevance, objectivity, and transparency). By providing ethical practice standards at this level of detail, the ASA Ethical Guidelines can usefully augment the professional's toolkit for ethical statistical practice.

---

[10] Tractenberg (2020, 2022-A, B) specifically evaluates the description of ethical practices of statistical practitioners, as reflected in the Ethical Guidelines for Statistical Practices maintained by the American Statistical Association (ASA, 2022) and the Association of Computing Machinery (ACM, 2018). Müller et al. (2021) explore codes of conduct (for mathematics practitioners), which discuss "the personal behavior of a professional in the field, enjoining such virtues as honesty, competence, respect between professionals and ensuring work does not bring the profession into disrepute" (p 40).

[11] The Evidence Act increases membership of ICSP to include the 24 federal agencies identified in the Chief Financial Officer Act. See 31 U.S.C. 901(b).

2. Examination across guidelines can identify strengths and create opportunities for growth

A better understanding of whether and how these guidance documents converge and/or differ has other practical benefits. At times of change, providing policy direction at institutional levels may lag while innovation in technical approaches occurs at the individual level. In other cases, guidelines may set principles at the institutional level that have not yet been developed as practices or ethics for the individual. Alignment across guidelines can "fill in" gaps or identify new areas for growth.

A better understanding of guidance documents can also identify areas of potential growth, either by identifying additional considerations, new branches of application, or the recognition of a need for greater clarity in guidance to promote uniform/uniformly ethical practice and inform decision making. Alignments across guidance documents may support or provide more complete and uniform guidance for users. Where alignment is lacking, opportunities for clarification or resolution may be found.

Examination across guidelines can also reinforce the federal guidelines when/if they come under attack. Historically, the federal principles have been considered largely uncontroversial, but recent events demonstrate that norms we have long taken for granted within the federal statistical system (e.g., independence from political interference) may not always be the case. Documenting the overlap among federal guidelines and those of the ASA Ethical Guidelines can help to buttress the importance of the federal guidelines when/if such conflicts arise again.

3. Identification can direct attention to resolve potential tensions across guidance and disciplines

Additionally, identifying and understanding differences (or tensions) in priorities, and the rationales behind these differences, encourages diverse perspectives to be taken into account when work crosses professions or is interdisciplinary. At minimum, this improves communication across partnerships, and supports ethical cross-boundary and interdisciplinary work. So too, apparent differences (or tensions) may be resolved by addressesing ambiguities or internal inconsistencies within and across frameworks at the routine and periodic revision to established guidelines.

4. Clearer understanding of the federal guidance resources can assist professional training

An understanding of the guidance for the profession (ASA) and for the federal context (US guidance), specifically, where they converge and where they diverge, can also be used for training of professional personnel. The analysis can illustrate the purposes and utility of different guidance resources, and underscore the value of the critical thinking needed when balancing (at times) competing priorities. In these cases, detailed guidance may be especially useful to identify where priorities do or could conflict, and how to address those conflicts in an ethical manner.

5. Questions an examination of guidelines cannot address

Of course, an examination of guidelines by itself cannot answer broader questions regarding the direction and implementation of statistical and data policy. It cannot provide a priority, or rank ordering, of goals to be achieved. Nor should such an examination of guidelines (and any differences observed) be taken out of the intended context by signaling and attributing unimportance of a given goal to a U.S. statistical body.

B. Data

This paper examines alignment across five key guidelines for US statistical and data science practice. As discussed throughout, these guidelines were developed at different times for different purposes and therefore different audiences. Long-standing guidances have been periodically reviewed and revised to ensure relevancy and clarity. Therefore, interpretation of results regarding areas of alignment, tension, or absence must take these differences into account.

**Box 1. Data sources examined by intended purpose and intended audience**

| Data Source | Updated | Intended Purpose | Intended Audience |
|---|---|---|---|
| ASA Ethical Guidelines (ASA EG) | 2022 | Code of professional ethics | Statistics practitioners |
| OMB Statistical Policy Directive 1 (SPD 1) | 2014 | Aspirational goals for federal statistical agencies | Federal statistical agencies (leadership) |
| OMB Statistical Policy Directive 2 and 2a (SPD2 and SPD2a) | 2006; 2015 | Aspirational practices for federal statistical agencies | Federal statistical agencies (program staff) |
| NASEM Principles and Practices (PNP) | 2021 | Aspirational goals and practices | Federal statistical agencies (leadership) |
| OMB Data Ethics Tenets (DET) | 2020 | Aspirational goals | Federal staffs engaged in data lifecycle |

In a separate paper, we use the same alignment method and model to examine congruence in guidelines from the ASA and diverse international organizational bodies (ASA Ethical Guidelines and the UN Fundamental Principles of Official Statistics, OECD Good Statistical Practice, and European Statistics Code of Practice).

1. ASA Ethical Guidelines for Statistical Practice

Founded in 1839, the American Statistical Association is one of the oldest continuously operating professional organizations for statistical practice in the world. Its mission is to support the practice and profession of statistics. The ASA has benefitted from an international membership since its inception.

The ASA Ethical Guidelines for Statistical Practice (ASA EG) were first published in 1982 as a three-year trial. They were motivated in part by a report published in 1973 outlining the results of the American Statistical Association-Federal Statistics Users' Conference Committee on the Integrity of Federal Statistics. Nonetheless, federal statisticians represent a small portion of the broader, intended audience of the ASA EG. Indeed, the ASA EG are intended as the ethical practice standard for all statistics and data science practitioners (Tractenberg, 2020; 2022-A, 2022-B). GIven the international reach of the ASA, the ASA EG are endorsed by professional associations in other countries.

Since 2016, the ASA EG have been reviewed every five years (Hogan & Steffey, 2014). The most recent revision of the ASA EG was approved in February 2022. In their current form, the ASA EG include 60 items under eight general areas, plus a new 12-item Appendix, specific to institutions and organizations, created during this update. Revisions were made to extend applicability to a wider range of practitioners and contexts, and to ensure that guidance regarding misconduct was concrete and actionable. The number of elements supporting each of the eight general areas is shown in parentheses in Box 2 and presented in detail in Annex 1.

**Box 2. ASA Ethical Guidelines for Statistical Practice**

| |
|---|
| A. Professional Integrity & Accountability: Professional integrity and accountability require taking responsibility for one's work. Ethical statistical practice supports valid and prudent decision making with appropriate methodology. The ethical statistical practitioner represents their capabilities and activities honestly, and treats others with respect. (12 elements) |
| B. Integrity of data and methods: The ethical statistical practitioner seeks to understand and mitigate known or suspected limitations, defects, or biases in the data or methods and communicates potential impacts on the interpretation, conclusions, recommendations, decisions, or other results of statistical practices. (7 elements) |
| C. Responsibilities to Stakeholders: Those who fund, contribute to, use, or are affected by statistical practices are considered stakeholders. The ethical statistical practitioner respects the interests of stakeholders while practicing in compliance with these Guidelines. (8 elements) |
| D. Responsibilities to research subjects, data subjects, or those directly affected by statistical practices: The ethical statistical practitioner does not misuse or condone the misuse of data. They protect and respect the rights and interests of human and animal subjects. These responsibilities extend to those who will be directly affected by statistical practices. (11 elements) |
| E. Responsibilities to members of multidisciplinary teams: Statistical practice is often conducted in teams made up of professionals with different professional standards. The statistical practitioner must know how to work ethically in this environment. (4 elements) |
| F. Responsibilities to Fellow Statistical Practitioners and the Profession: Statistical practices occur in a wide range of contexts. Irrespective of job title and training, those who practice statistics have a responsibility to treat statistical practitioners, and the profession, with respect. Responsibilities to other practitioners and the profession include honest communication and engagement that can strengthen the work of others and the profession. (5 elements) |
| G. Responsibilities of Leaders, Supervisors, and Mentors in Statistical Practice: Statistical practitioners leading, supervising, and/or mentoring people in statistical practice have specific obligations to follow and promote these Ethical Guidelines. Their support for – and insistence on – ethical statistical practice are essential for the integrity of the practice and profession of statistics as well as the practitioners themselves. (5 elements) |
| H. Responsibilities regarding potential misconduct: The ethical statistical practitioner understands that questions may arise concerning potential misconduct related to statistical, scientific, or professional practice. At times, a practitioner may accuse someone of misconduct, or be accused by others. At other times, a practitioner may be involved in the investigation of others' behavior. Allegations of misconduct may arise within different institutions with different standards and potentially different outcomes. The elements that follow relate specifically to allegations of statistical, scientific, and professional misconduct. (8 elements) |
| APPENDIX: Responsibilities of organizations/institutions: Whenever organizations and institutions design the collection of, summarize, process, analyze, interpret, or present, data; or develop and/or deploy models or algorithms, they have responsibilities to use statistical practice in ways that are consistent with these Guidelines, as well as promote ethical statistical practice. (Organizations 7 elements; Leaders 5 elements; 12 elements total) |

Although organizations are discussed in the ASA EG Appendix, they are grouped together into a single column with ASA Principle G in our alignment tables because of the potential overlap of institutions that are not statistical practitioners (guidance located in Appendix) and leaders, supervisors, and mentors who are statistical practitioners (guidance located in Principle G).

## 2. OMB Statistical Policy Directive 1

The U.S. Office of Management and Budget has had, essentially since the 1930's, a central role in establishing statistical policy and standards. Currently, through the Office of Information and Regulatory Affairs, the U.S. Chief Statistician has responsibility for the coordination of the federal statistical system under the authority of the Budget and Accounting Procedures Act of 1950 (31 U.S.C. 1104(d)) and the Paperwork Reduction Act of 1995 (44 U.S.C. 3504 (e)). There are thirteen principal federal statistical agencies whose mission is to produce official statistics; another three recognized statistical units share that mission within their respective program offices.[12] Additionally, there are more than a hundred other federal agencies whose core mission is not the production of statistics engaged in substantial statistical activities. (*Statistical Programs of the United States Government, 2019/20*). The passage of the Foundations for Evidence-Based Policy Act in 2018, and the subsequent guidance issued by OMB M-19-23, have increased opportunities–and expectations– for the U.S. Chief Statistician to engage with the statistical programs in CFO agencies, particularly in the areas of data quality, confidentiality, access, and governance.[13]

---

[12] The U.S. principal federal statistical agencies are (13): Bureau of Economic Analysis (Department of Commerce); Bureau of Justice Statistics (Department of Justice); Bureau of Labor Statistics (Department of Labor); Bureau of Transportation Statistics (Department of Transportation); Census Bureau (Department of Commerce); Economic Research Service (Department of Agriculture); Energy Information Agency (Department of Energy); National Agricultural Statistics Service (Department of Agriculture); National Center for Education Statistics (Department of Education); National Center for Health Statistics (Department of Health and Human Services); National Center for Science and Engineering Statistics (National Science Foundation); Office of Research, Evaluation, and Statistics (Social Security Administration); and Statistics of Income (Department of Treasury). There are also three recognized federal statistical units: Microeconomic Surveys Unit (Federal Reserve Board); Center for Behavioral Health Statistics and Quality (Substance Abuse and Mental Health Services Administration; Department of Health and Human Services); and National Animal Health Monitoring System (Animal and Plant Health Inspection Service, Department of Agriculture).

[13] "The statutory purpose of the Interagency Council on Statistical Policy (ICSP) is to advise and assist the Chief Statistician of the United States in the coordination of the Federal Statistical System; the implementation of statistical policies, principles, standards, and guidelines; and the evaluation of statistical program performance.[Prior to the passage of the Evidence Act,] Its membership included representatives of 14 member agencies. [With the passage of the Evidence Act,] these member agency representatives are joined by all statistical officials designated pursuant to section 2(b) of this guidance that are not already members, increasing ICSP membership by 12. Consequently, the ICSP will include two members each from the Department of Commerce and the USDA, which each contain two principal statistical agencies, and one member each from all other CFO Act agencies. The expanded membership will augment the ICSP's ability to meet its statutory purpose. In addition, it will be a forum for collaboration, coordination, and information sharing among the statistical agencies and additional statistical programs across all member agencies, including on issues such as ensuring data quality and confidentiality, attaining and providing data access, and playing an effective role in agency-wide data governance." (p. 9, M-19-23)

Federal statistical policy directives are one mechanism through which coordination of the federal statistical system occurs. OMB has issued several statistical policy directives over the decades to guide the production of comparable information for federal statistical purposes. They are developed through interagency collaboration and public comment, and generally represent minimum guidelines. As federal regulations, statistical policy directives take a form that is intended to be easily communicated. Implementation guidance is often issued to describe practices necessary to implement the directive.

Most of these directives pertain to specialized areas of measurement, such as occupation ([Statistical Policy Directive 10](#)) and industries ([Statistical Policy Directive 8](#)); metropolitan statistical areas ([Statistical Policy Directive 7](#)); race/ethnicity ([Statistical Policy Directive 15](#)); and poverty ([Statistical Policy Directive 14](#)). A select few of statistical policy directives are broader in their scope, providing guidance across federal statistical agencies and their programs–for example, the collection, release, and dissemination of principal federal economic indicators ([Statistical Policy Directive 3](#)) and the release and dissemination of statistical products produced by federal statistical agencies ([Statistical Policy Directive 4](#)).

However, two statistical policy directives are truly foundational to the goals and practice of federal statistical agencies, one of which was codified in law, and therefore are the focus of the examinations in this paper. In 2014, OMB issued [Statistical Policy Directive 1](#) (SPD1): *Fundamental Responsibilities of Federal Statistical Agencies and Recognized Statistical Units*. This directive articulated long-held and commonly followed expectations for federal statistical agencies and designated statistical units. In many cases, further guidance regarding the achievement of these responsibilities can be found in federal law, other statistical policy directives, and federal agency policy. The issuance of SPD1 was intended to improve communication of these core responsibilities among policy stakeholders rather than offer guidance for practitioners; likewise, it was not intended to propose new or expand existing requirements. See Box 3. In 2018, Statistical Policy Directive 1 was embedded within section 3563 the [Foundations for Evidence-Based Policy Act](#).

**Box 3. OMB Statistical Policy Directive 1**

| | |
|---|---|
| 1 | Produce and disseminate relevant and timely information |
| 2 | Conduct credible and accurate statistical activities |
| 3 | Conduct objective statistical activities |
| 4 | Protect the trust of information providers by ensuring the confidentiality and exclusive statistical use of their responses |

[Statistical Policy Directive 2](#): *Standards and Guidelines for Statistical Surveys* was issued in 2006 to provide guidance to federal statistical programs submitting information clearance requests for OMB review and approval under the Paperwork Reduction Act.[14] The directive describes expectations for the conduct of federal agencies throughout the information collection lifecycle. The form of the directive is intended to be sufficiently broad to cover differing scope and methods of information collections used across the full federal statistical system, yet detailed enough to support evaluation of the expected quality and burden associated with the proposed information collection. In 2015, an addendum to Statistical Policy Directive 2 was issued: Standards and Guidelines for Cognitive Interviews. Here, the addendum is referred to as Statistical Policy Directive 2a.

---

[14] In fact, the 2006 issuance was an update of prior methodological directives 1 and 2 that had been in use for several decades as attachments to the original OMB Circular A-46, which set forth the series of standards as attachments. During the 1970s, these standards became known as Directives when the Statistical Policy Office resided in the Commerce Department, and renamed Statistical Policy Directives when the Office returned to the Office of Management and Budget under the Paperwork Reduction Act (Katherine Wallman, personal communication, April 2023).

**Box 4. OMB Statistical Policy Directives 2 and 2a**

| | |
|---|---|
| **SPD 2: Standards and Guidelines for Statistical Surveys** | |
| **1** | ***Development of concepts, methods, and design*** |
| 1.1 | Survey planning |
| 1.2 | Survey design |
| 1.3 | Survey response rates |
| 1.4 | Pretesting survey systems |
| **2** | ***Collection of data*** |
| 2.1 | Developing sampling frames |
| 2.2 | Required notification of potential survey respondents |
| 2.3 | Data collection methodology |
| **3** | ***Processing and editing of data*** |
| 3.1 | Data editing |
| 3.2 | Nonresponse analysis and response rate calculation |
| 3.3 | Coding |
| 3.4 | Data protection |
| 3.5 | Evaluation |
| **4** | ***Protect the trust of information providers*** |
| 4.1 | Developing estimates and projections |
| **5** | ***Data analysis*** |
| 5.1 | Analysis and report planning |
| 5.2 | Inference and comparisons |
| **6** | ***Review procedures*** |
| 6.1 | Review of information products |
| **7** | ***Dissemination of information products*** |
| 7.1 | Releasing information |
| 7.2 | Data protection and disclosure avoidance for dissemination |
| 7.3 | Survey documentation |
| 7.4 | Documentation and release of public use microdata |
| **SPD 2a: Standard and Guidelines for Cognitive Interviews** | |
| A.1 | Methodological plan |
| A.2 | Sample selection |
| A.3 | Interview guide |
| A.4 | Analysis of cognitive interviews |
| A.5 | Transparent analysis |
| A.6 | Final reports |
| A.7 | Reporting results |

### 3. NASEM Principles and Practices for Federal Statistical Agencies

Under the aegis of the National Academies of Sciences, Engineering, and Medicine, the Committee on National Statistics (CNSTAT) provides "…advice to the federal government and the nation grounded in the current best scientific knowledge and practice that will lead to improved statistical methods and information upon which to base public policy." Since its inception in 1973, CNSTAT has produced roughly 300 reports on federal statistical programs, surveys, and statistical methods. (NASEM, 2021 p. 21)[15]

As part of its mission to provide an independent review of federal statistical activities, CNSTAT issues its report, *Principles and Practices for Federal Statistical Agencies (NASEM PNP)*, as a way to communicate to policymakers "characteristics of statistical agencies that enable them to serve the public good." (NASEM, 2021, p. 5) The first edition was issued in 1992. The principles describe aspirational goals for federal statistical agencies; practices describe means to achieve these goals. Rather than a blueprint for aligning particular practices to particular principles, generally several principles and practices are intended to be achieved when jointly pursued. Although principal federal statistical agencies and designated units are the primary intended audience, PNP guidelines are widely used across federal agencies conducting statistical work and other statistics practitioners in the U.S.

Since its second edition in 2001, the report has been reviewed and, as necessary, revised over time to reflect changes to national data infrastructure, as well as legal and regulatory frameworks, and thereby maintain its relevance. These editions are distributed every four years to coincide with the beginning of each presidential administration or second term.

Compared to the 2017 report, the current edition expands the list of principles from four to five by elevating a previously underlying theme--continual improvement and innovation--to the status of a principle. The report also consolidates some practices that, previously identified separately, are closely entwined. (NASEM, 2021 p. 22-23)

**Box 4. NASEM Principles and Practices for Federal Statistical Agencies**

| |
|---|
| Principle 1. Relevance to policy issues and society |
| Principle 2. Credibility among data users and stakeholders |
| Principle 3. Trust among the public and data providers |
| Principle 4. Independence from political and other undue external influence |
| Principle 5. Continual improvement and innovation |
| Practice 1. A clearly defined and well accepted mission |
| Practice 2. Necessary authority and procedures to protect independence |
| Practice 3. Commitment to quality and professional standards of practice |
| Practice 4. Professional advancement of staff |
| Practice 5. An active research program |
| Practice 6. Strong internal and external evaluation processes for an agency's statistical programs |
| Practice 7. Coordination and collaboration with other statistical agencies |
| Practice 8. Respect for data providers |
| Practice 9. Dissemination of statistical products that meet users' needs |
| Practice 10. Openness about sources and limitations of the data provided |

4. OMB Data Ethics Tenets

The U.S. Federal Data Strategy was issued in 2019 to advance access, interoperability, and utility of federal data. Its 2020 Action Plan called for the development of a Data Ethics Framework to help agency employees, managers, and leaders make ethical decisions as they acquire, manage, and use data throughout the data lifecycle. The intended audience for this framework is all federal agencies involved in the data lifecycle–extending beyond federal statistical agencies and statistical programs. To that end, an interagency group was formed by the General Services Administration (GSA) comprising federal government agencies with expertise in statistics, public policy, evidence-based decision making, privacy, and analytics. Input was also received from the Chief Data Officer (CDO) Council, Interagency Council on Statistical Policy (ICSP), and the Federal Privacy Council (FPC).

---

[15] National Academies of Sciences, Engineering, and Medicine. (2021). Principles and Practices for a Federal Statistical Agency, Seventh Edition. Washington, DC: The National Academies Press. https://doi.org/10.17226/25885.

The resulting Framework was intended to be a "living" resource and to be regularly updated by the CDO Council and ICSP. As described in its Framework document, "data ethics are the Data Ethics Framework norms of behavior that promote appropriate judgments and accountability when collecting, managing, or using data, with the goals of protecting civil liberties, minimizing risks to individuals and society, and maximizing the public good." (p. 5, OMB Data Ethics Tenets 2020)

**Box 5. OMB Data Ethics Tenets**

| 1 | Uphold applicable statutes, regulations, professional practices, and ethical standards |
|---|---|
| 2 | Respect the public, individuals, and communities |
| 3 | Respect privacy and confidentiality |
| 4 | Act with honesty, integrity, and humility |
| 5 | Hold oneself and others accountable |
| 6 | Promote transparency |
| 7 | Stay informed of developments in the fields of data management and data science |

C. Methods
1. Degrees of Freedom Analysis

Examining qualitative data, such as professional guidance, in a systematic, rigorous, but easily communicated manner is challenging. The Degrees of Freedom Analysis (DoFA, Campbell 1975; Tractenberg 2023) approach is a structured approach designed to facilitate the combination of qualitative information into a coherent, interpretable, analysis framework. It is a form of qualitative content analysis where the investigator searches for patterns and themes in narrative (Nieswiadomy & Bailey, 2018). Originally intended for theory building (Campbell 1975), this method was adapted for decision making by Tractenberg (2019) and was applied to the analysis of alignment of ethical practice standards across diverse disciplines in Rios et al. (2019) and Tractenberg (2020). As outlined in Tractenberg (2019, 2023) and demonstrated in Tractenberg (2020, 2022-A, 2022-B), the matrix of the evidence or information to be studied is assembled into a single two-dimensional table.

For this application, the analysis is focused on the identification of agreement between elements of two different guidance documents (e.g., ASA EG and NASEM PNP), i.e., "alignment". Quantitation (i.e., the calculation of actual degrees of freedom) is not necessary in this two-dimensional alignment (Tractenberg 2019; Tractenberg 2020). The method was used extensively in Tractenberg 2022-A, Tractenberg 2022-B and is more fully described in Tractenberg 2023.

2. Code assignment

As a first step, the first author examined each guidance pair and applied detailed codes. Once the coding was completed, the second author evaluated the results to discuss the contents or coding of all cells in that table. If the two authors disagreed on any match (schema described below), further discussion determined if consensus could be reached, or if that result should be identified as an area where further clarity would improve the interpretation and use of guidance documents. Detailed coded tables are presented in Annex 2.

Subsequently, the detailed coded tables were summarized into themes illustrated using a simplified coding strategy to improve communication of results; these summary tables are presented in the body of the report. (See below.) Note, although it is possible that different coders may generate somewhat different detailed codes than those prepared by the authors,[16] we are confident that the pattern of themes summarized in this paper would be consistent, assuming familiarity with the guidelines examined.

3. Coding schema

Recall (as described in the section above), the guidance documents examined for this analysis were developed for different purposes and different audiences. Thus, observations regarding alignment, tensions, and potential conflict should take these differences into account in coding schemes and interpretation of results. Accordingly, we adjusted our coding schema to capture these nuances.

For this analysis, we determined the coding scheme to describe alignment, tension, or potential conflict across each pair of guidelines examined *a priori;* the results are summarized in two-dimensional tables. To generate a consistent and interpretable "signal", all coding is in terms of the ASA EG. In other words, coding identifies specific elements of the ASA EG that align with (i.e., agree with, support) each of the elements from the other examined guidance document. Areas of tension, and potential conflict were also identified. The next section provides more detail regarding specific coding schema.

Box 6 provides a key to our coding schema. Definitions of terms and examples from our observations of guideline pairs are provided alongside exemplar detailed codes and summary codes. For example, a particular observation in the examination of a guidance document pair could indicate a scenario in which similar ideas, expressed as goals, products, or behavior, will be supportive of one another across guidance documents. These "alignments" therefore may be literal or similar matches (see (a) and (b), below). If a pattern of support is found, the relevant row or column is shaded in green (see (f) below).

In other cases, a particular observation could indicate a scenario in which certain ideas (again, expressed as goals, products, or behavior) across guidance documents are opposed (see (c) below). These "conflicts" may appear alone, or co-exist with areas of alignment so the resulting observation describes "tension" (see (d) below). If a pattern of tension (conflict and alignment) is found, the relevant row or column is shaded in yellow (see (g) below).

There are also potential scenarios of apparent "guidance gaps" within a particular guidance document that are represented in the other guidance document for a given pair. For a given instance, this is shown by a blank cell (see (e) below). Patterns of this scenario are shown by shading the relevant row or column in gray (see (h) below).

**Box 6. Key to alignment model coding schema**

| Scenario | Example observation | Example detail code | Example summary code |
|---|---|---|---|

---

[16] See Tractenberg and Gordon (2017) for a fuller discussion of DoFA and its reliability and validity across content types/narratives.

| a | Literal or conceptual match | If there is a thematic or exact match to the particular guideline, the particular ASA EG element is indicated | D[17] | ✓ [check] |
|---|---|---|---|---|
| b | Similar match | If the alignment with a particular guideline is abstract, or held under only specific circumstances, then the particular ASA EG element is indicated in parenthesis | (D)[18] | (✓) [check] |
| c | Opposition | If a particular guideline could potentially cause conflict, and specific attention is required for balanced implementation of guidance, then the ASA EG element appears in red font with a ~ ("not") | ~D10[19] | ~ [not] |
| d | Tension | If both alignment and potential conflict coexist for a particular guideline, then both ASA EG aligned and opposed elements are indicated | A2,~A3, ~A4[20] | ✓ [check], ~ [not] |
| e | Unmatched | If there is neither exact nor thematic alignment (or nonalignment) | [blank][21] | [blank] |
| f | Complete alignment | Patterns of complete alignment across that row (or column) | D | ✓ [check], (✓) [check] |
| g | Complete tension | Patterns of alignment AND tension are identified across that row (or column) | B, ~D, | ✓ [check], ~ [not] |
| h | Guidance gap | Patterns of unmatched content are identified across that row (or column) | [blank] | [blank] |

---

[17] For example, we observe strong alignment between ASA EG D4: *"(the ethical statistical practitioner) Protects people's privacy and the confidentiality of data concerning them, whether obtained from the individuals directly, other persons, or existing records. Knows and adheres to applicable rules, consents, and guidelines to protect private information."* and OMB SPD 1 Responsibility 4, *"Protect the trust of information providers by ensuring the confidentiality and exclusive statistical use of their responses."*

[18] For example, we observe an inexact alignment when examining the detailed description of guidance provided by ASA EG D: "Research Subjects/Data Subjects and Those Affected by Statistical Practices" and OMB SPD 2, Survey Planning Standard 1.1: Develop a written plan.

[19] For example, we observe a potential conflict between ASA EG D10, *"(the ethical statistical practitioner) Understands the provenance of the data, including origins, revisions, and any restrictions on usage, and fitness for use prior to conducting statistical practices"* and in achieving NASEM PNP Principle 1, "Relevance to Policy Issues and Society" if data unfit for purpose, or for which permissions have not been obtained, are deemed to be "relevant" for a policy issue.

[20] OMB SPD 1 Responsibility 1, *"Produce and disseminate relevant and timely information"*, is well-aligned with ASA EG A2. *"Uses methodology and data that are valid, relevant, and appropriate, without favoritism or prejudice, and in a manner intended to produce valid, interpretable, and reproducible results,"* but could potentially cause conflict with achievement of ASA EG A3 *"Does not knowingly conduct statistical practices that exploit vulnerable populations or create or perpetuate unfair outcomes."* Therefore, we observe potential tensions that could require balancing approaches to ensure the achievement of both guidelines.

[21] For example, ASA EG Principle F, responsibilities relating to other statistical practitioners and the profession offers no support or guidance for how to accomplish NASEM PNP Practice 9, "Dissemination of Statistical Products That Meet Users' Needs."

## D. Results

See Tables 1-3 for a summary of themes alignment between ASA EG, OMB SPD 1, NASEM PNP, and OMB DET, respectively. Tables 4-6 show the alignment among these latter three guidelines. Analog detail tables are presented in Annex 2.

### 1. Alignment of ASA Ethical Guidelines with OMB Statistical Policy Directives 1, 2, and 2a

*Target Audiences*: Recall that the OMB SPD were developed to communicate high-level expectations of leadership within federal statistical agencies and their programs. Accordingly, SPD are intended to pertain to a select portion of the professional statistician and data science community.

*Areas of Strong Alignment*: As shown in Table 1[22], there was complete alignment of ASA EG with OMB SPD1 Responsibilities 2 (*credible*) and 3 (*objective*). Both of these responsibilities were aligned with all eight ASA EG principles; this pattern is denoted by green shading. Further, in many cases, the particular practices and behaviors articulated by the ASA EG directly support achievement of SPD1 Responsibilities 2 and 3. With regard to OMB SPD1 Responsibility 4 (*trust*), we noted some alignment from ASA EG Principles C (stakeholders), and, particularly, Principle D (data providers) (see Annex 2, Detail Table 1).

Turning to OMB SPD 2 and 2a, we see strong alignment with ASA EG Principles A (accountability), B (integrity), and C (stakeholders). In addition, OMB SPD 2 also shows strong alignment with ASA EG Principle D (data providers) ; this pattern is denoted by green shading.

| ASA Ethical Guidelines → <br> OMB SPD Guidelines ↓ | A Accountability | B Integrity | C Stakeholder | D Data Providers | E Other Disciplines | F Other Statisticians | G Leadership, Appendix | H Misconduct |
|---|---|---|---|---|---|---|---|---|
| 1.1 Relevant | ✓ ~ | ✓ | ✓ | ✓ ~ | ✓ | | ✓ ~ | ✓ |
| 1.2 Credible and accurate | ✓ | ✓ | ✓ | ✓ | ✓ | ✓ | ✓ | ✓ |
| 1.3 Objective | ✓ | ✓ | ✓ | ✓ | ✓ | ✓ | ✓ | ✓ |
| 1.4 Trust | | | ✓ | ✓ | | | | |
| 2.1.1 Written plan | ✓ | ✓ | ✓ | ✓ | ✓ | | ✓ | |
| 2.1.2 Survey design | ✓ | ✓ | ✓ | ✓ | ✓ | | ✓ | |
| 2.1.3 Response rate design | ✓ | ✓ | ✓ | ✓ | | | ✓ | |
| 2.1.4 Functioning components | ✓ | ✓ | ✓ | ✓ | ✓ | | ✓ | |
| 2.2.1 Appropriate frame | ✓ | ✓ | ✓ | ✓ | | | | |
| 2.2.2 Notify respondents | ✓ | ✓ | ✓ | ✓ | ✓ | | ✓ | |
| 2.2.3 Balance quality v burden | ✓ | ✓ | ✓ | ✓ | ✓ | | | |
| 2.3.1 Appropriate data edits | ✓ | ✓ | ✓ | ✓ | ✓ | | ✓ | ✓ |
| 2.3.2 Nonresponse analysis | ✓ | ✓ | ✓ | ✓ | | | ✓ | |
| 2.3.3 Quality for other study | ✓ | ✓ | ✓ | ✓ | | ✓ | ✓ | |
| 2.3.4 Avoid disclosure | ✓ | ✓ | ✓ | ✓ | ✓ | | ✓ | ✓ |
| 2.3.5 Evaluate data quality | ✓ | ✓ | ✓ | ✓ | ✓ | ✓ | ✓ | |
| 2.4.1 Use theory and methods | ✓ | ✓ | ✓ | ✓ | ✓ | | ✓ | ✓ |
| 2.5.1 Analysis plan | ✓ | ✓ | ✓ | ✓ | ✓ | | | |
| 2.5.2 Good statistical practice | ✓ | ✓ | ✓ | ✓ | ✓ | | ✓ | |
| 2.6.1 Review dissemination | ✓ | ✓ | ✓ | ✓ | ✓ | ✓ | ✓ | |
| 2.7.1 Equitable dissemination | ✓ | ✓ | ✓ | ✓ | | | ✓ | |
| 2.7.2 Data protection | ✓ | ✓ | ✓ | ✓ | | | | |
| 2.7.3 Survey documentation | ✓ | ✓ | ✓ | ✓,~ | ✓ | ✓ | | |
| 2.7.4 Public use microdata | ✓ | ✓ | ✓ | ✓ | ✓ | ✓ | ✓ | |
| 2.A.1 Methodological plan | ✓ | ✓ | ✓ | ✓ | ✓ | ✓ | ✓ | |
| 2.A.2 Sample selection standard | ✓ | ✓ | ✓ | ✓,~ | | | | |
| 2.A.3 Interview guide standard | ✓ | ✓ | ✓ | | | | | |
| 2.A.4 Systematic analysis | ✓ | ✓ | ✓ | ✓ | ✓ | ✓ | ✓ | |
| 2.A.5 Transparent analysis | ✓ | ✓ | ✓ | ✓ | | | | |
| 2.A.6 Final reports standard | ✓ | ✓ | ✓ | ✓ | ✓ | | | |
| 2.A.7 Reporting results standard | ✓ | ✓ | ✓ | ~ | ✓ | ✓ | ✓ | |

*Table 1 Summary of themes identified in ASA Ethical Guidelines and OMB Statistical Policy Directives 1, 2, 2a*

*Areas of Tension*: With regard to OMB SPD 1, we observe potential tensions between achievement Responsibility 1 (*relevance*) and implementation of ASA EG. Across six of the ASA EG principles, achievement of Responsibility 1 was supported in most cases by multiple elements within each ASA EG principle. However, achievement of Responsibility 1 could cause tension with implementation of certain ASA EG. For example, ASA EG Principle A (accountability), which describes professional expectations for statistical product accuracy, could impede timely release. Similarly, implementation of the ASA EG Principle D (data providers) describing expectations for protection of data providers could impede release of relevant statistics if doing so might cause harm. In these cases, particular attention should be paid in the implementation of OMB SPD 1.

---

[22] Key: Areas of alignment are indicated by ✓ (check). Areas of conflict are indicated by ~ (not). Areas of tension (both alignment and potential conflict) are indicated by ✓ (check) and ~ (not). Patterns of alignment are shaded in green. Patterns of tension are shaded in yellow. Patterns of gaps in guidance are shaded in gray.

Turning to OMB SPD 2 and 2a, we observe potential tensions with regard to ASA EG Principle D (data providers) with regard to 2.7.3 (survey documentation), and 2.A.2 (sample selection standard). We also note a potential conflict with regard to 2.A.7 (reporting results standard). As is shown in the detailed table (Appendix), ensuring that reporting is carried out in a standard fashion might lead to conflicts with ASA EG elements that seek to protect vulnerable populations, and prompt practitioners to consider the potential impacts of their work on groups and subgroups. Transparency and methodological rigor (including reporting standards) are prioritized throughout the EGs, but data contributors' rights must also be taken into consideration. Thus, tension is identified.

*Guidance gap*: We observe that SPD 1 Responsibility 4 (trust) could benefit from including further guidance on practices and behaviors, such as found in the ASA EG.

*Discussion:* Overall, our method shows generally strong alignment between the ASA EG and SPD 1, 2, and 2a. Although the intended audiences for each guidance are similar, SPD generally are developed for a targeted group within the overall professional statistician and data science community. In some areas, alignment is particularly strong–such as guidance regarding *credible* and *objective* statistics. These areas of strong alignment across guidance documents provide a particularly solid basis for strengthening existing guidance through higher level authority, such as regulation or law.

However, we also observed areas of tension across guidance documents–notably, *relevance*. We also observed a guidance gap in the area of *trust*. Upon examination, we found these tensions could be addressed through clarification of language when implementing routine reviews and possible revision of OMB SPD 1, 2 and 2a. Because the ASA EG describe professional expectations for how practices should be implemented (or behaviors), they may be particularly useful to such reviews. Areas of tension should be resolved prior to elevating guidance to higher levels of authority, such as regulation or law.

2. Alignment of ASA Ethical Guidelines with National Academies' Principles and Practices

*Target Audiences:* Recall that the NASEM PNP were developed to communicate high-level expectations of leadership in federal statistical agencies. Accordingly, NASEM PNP are intended to pertain to a select portion of the professional statistician and data science community.

*Areas of Strong Alignment*: As shown in Table 2[23], ASA EG supports the accomplishment of each of the NASEM PNP overall. There was complete alignment of all eight ASA EG principles with NASEM Principles 2 (*credibility)*, 3 (*trust)*, and 4 (*independence)*. There was also complete alignment of ASA EG with NASEM Practices 2 (a*uthority)* and 3 (*commitment)*. In addition, there was notable alignment across the ASA EG with NASEM Principles 1 (*relevance) and 5*

---

[23] Key: Areas of alignment are indicated by ✓ (check). Areas of conflict are indicated by ~ (not). Areas of tension (both alignment and potential conflict) are indicated by ✓ (check)and ~ (not). Patterns of alignment are shaded in green. Patterns of tension are shaded in yellow. Patterns of gaps in guidance are shaded in gray.

(*improvement*); and NASEM Practices 1 (*mission*), 4 (*staff*), 6 (*evaluation*), 8 (*respect*) (supported by every ASA Principle D element, shaded in green), and 10 (*transparency*).

*Areas of Tension:* In some cases, however, the principles, practices, and behaviors described in NASEM PNP, while largely complementary, can potentially lead to tension. We observed scenarios where the implementation of NASEM PNP could potentially conflict with adherence to each of the eight ASA EG principles. This tension was particularly notable when examining NASEM Practices 7 (*collaboration*) and 9 (*users' needs*) because the implementation–without caution–could conflict with elements of ASA EG Principles A (*accountability*), D (*data providers*), and G (*supervisors/mentors*).

*Guidance Gaps*: We observe that NASEM Principle 5 and Practices 4, and 10 (*improvement, staff,* and *transparency*, respectively) could benefit from including further guidance on practices and behaviors, such as found in the ASA EG.

*Discussion:* Overall, our method shows generally strong alignment between the ASA EG and NASEM PNP. As with OMB SPD, although the intended audiences for each guidance are similar, NASEM PNP are developed for a targeted group within the overall professional statistician and data science community. In some areas, alignment is particularly strong: guidance regarding *credibility, trust, independence, authority, and commitment*. These areas of strong alignment across guidance documents provide a particularly solid basis for strengthening existing guidance through higher level authority, such as regulation or law.

| NASEM P and P (2021): \ ASA Ethical Guidelines (2022): | A Accountability | B Integrity | C Stakeholder | D Data Providers | E Other Disciplines | F Other Statisticians | G Leadership, Appendix | H Misconduct |
|---|---|---|---|---|---|---|---|---|
| PRI1 Relevance | ✓~ | ✓ | ✓ | ✓~ | ✓ | | ✓~ | |
| PRI2 Credibility | ✓ | ✓ | ✓ | ✓ | ✓ | ✓ | ✓ | ✓ |
| PRI3 Trust | ✓ | ✓ | ✓ | ✓ | ✓ | ✓ | ✓ | ✓ |
| PRI4 Independence | ✓ | ✓ | ✓ | ✓ | ✓ | ✓ | ✓ | ✓ |
| PRI5: Improvement | ✓ | ✓ | ✓ | ~ | | | ✓ | |
| PRA1: Mission | ✓ | ✓ | ✓ | ✓ | ✓ | | ✓ | |
| PRA2: Authority | ✓ | ✓ | ✓ | ✓ | ✓ | ✓ | ✓ | ✓ |
| PRA3: Commitment | ✓ | ✓ | ✓ | ✓ | ✓ | ✓ | ✓ | ✓ |
| PRA6: Evaluation | ✓~ | ✓ | | ✓ | ✓ | ✓ | ✓ | ✓ |
| PRA7: Collaboration | ✓~ | ~ | ✓~ | ~ | ~ | ✓~ | ✓~ | ~ |
| PRA8: Respect | ✓ | ✓ | ✓ | ✓ | | ✓ | ✓ | |
| PRA9: Users' Needs | ~ | ✓~ | ✓~ | ✓~ | ✓~ | | ~ | ~ |
| PRA10: Transparency | ✓ | ✓ | ✓ | ✓ | | | ✓ | ✓ |

*Table 2 Summary of themes identified in ASA Ethical Guidelines and NASEM Principles and Practices*

However, we also observed areas of tension across guidance documents–notably, *relevance, collaboration, and users' needs*. We also observed a guidance gap in the area of *improvement, staff, and transparency*. Upon examination, we found these tensions could be addressed through clarification of language when implementing routine reviews and possible revision of NASEM PNP. Because the ASA EG describe professional expectations for how practices should be implemented (or behaviors), they may be particularly useful to such reviews. Areas of tension should be resolved prior to elevating guidance to higher levels of authority, such as regulation or law.

3. Alignment of ASA Ethical Guidelines with OMB Data Ethics Tenets

*Target Audiences:* Recall that the OMB DET were developed to communicate high-level expectations for leadership and relations with stakeholders across federal agencies involved in the data lifecycle–extending beyond federal statistical agencies and statistical programs. Accordingly, OMB DET pertain to a select portion of the professional statistician and data science community, but a broader audience than targeted by OMB SPD and NASEM PNP.

*Areas of Strong Alignment*: See Table 3.[24] Overall, three of the seven OMB DET are supported by every ASA EG. Specifically, Tenets 1 *(laws)*, 2 (*respect public*), and 4 (*integrity)* are fully supported by the ASA EG.

*Areas of Tension*: None observed.

*Guidance Gaps:* We observe that OMB DET Tenets 3 (*respect privacy*) and 7 (s*tay informed*) could benefit from including further guidance on practices and behaviors, such as found in the ASA EG.

*Discussion:* Overall, our method shows very strong alignment between the ASA

| ASA Ethical Guidelines (2022): \ Data Ethics Tenets (2020): | A Accountability | B Integrity | C Stakeholder | D Data Providers | E Other Disciplines | F Other Statisticians | G Leadership | H Misconduct |
|---|---|---|---|---|---|---|---|---|
| 1 Uphold laws and regulations | ✓ | ✓ | ✓ | ✓ | ✓ | ✓ | ✓ | ✓ |
| 2 Respect public | ✓ | ✓ | ✓ | ✓ | ✓ | ✓ | ✓ | ✓ |
| 3 Respect privacy and confidentiality | ✓ | ✓ | ✓ | ✓ | ✓ | | | |
| 4 Integrity | ✓ | ✓ | ✓ | ✓ | ✓ | ✓ | ✓ | ✓ |
| 5 Accountability | ✓ | | ✓ | ✓ | ✓ | ✓ | ✓ | ✓ |
| 6 Transparency | ✓ | ✓ | ✓ | | ✓ | ✓ | ✓ | ✓ |
| 7 Stay informed | | ✓ | ✓ | | | ✓ | ✓ | |

*Table 3 Summary of themes identified in ASA Ethical Guidelines and OMB Data Ethics Tenets*

EG and OMB DET. As with OMB SPD and NASEM PNP, although the intended audiences for each guidance are similar, OMB DET are developed for a targeted group within the overall professional statistician and data science community. Alignment is particularly strong in guidance areas: *uphold law, respect the public*, and *integrity*. These areas of strong alignment across guidance documents provide a particularly solid basis for strengthening existing guidance through higher level authority, such as regulation or law.

Although we observed no areas of tension across guidance documents, we found guidance gaps in the areas of *respect privacy and confidentiality* and *stay informed*. These tensions could be addressed through clarification of language when implementing routine reviews and possible revision of OMB DET. Because the ASA EG describe professional expectations for how practices should be implemented (or behaviors), they may be particularly useful to such reviews. Areas of tension should be resolved prior to elevating guidance to higher levels of authority, such as regulation or law.

4. Overall alignment among examined federal guidelines

To better understand overall alignment, tension, and guidance gaps across all four guideline documents, we continued our examination of the alignment between pairs of national guidelines.

a. OMB Statistical Policy Directives 1, 2, and 2a and NASEM Principles and Practices

*Target Audiences:* Recall that both the OMB SPD and NASEM PNP were developed to communicate high-level expectations of leadership within federal statistical agencies and their programs. Accordingly, their target audiences are very similar–unlike the target audience for the ASA EG and, to some extent, OMB DET.

---

[24] Key: Areas of alignment are indicated by ✓ (check). Areas of conflict are indicated by ~ (not). Areas of tension (both alignment and potential conflict) are indicated by ✓ (check) and ~ (not). Patterns of alignment are shaded in green. Patterns of tension are shaded in yellow. Patterns of gaps in guidance are shaded in gray.

*Areas of Strong Alignment:* See Table a. [25] All responsibilities of SPD 1 are supported by several elements of NASEM PNP. However, alignment is more consistent among principles than among practices, particularly across OMB SPD 1 Responsibility 2 (*credibility*) and Responsibility 3 (*objectivity*). The strongest alignments among OMB SPD 1, 2, and 2a were observed among NASEM Principles 2 and 3 (*credibility, trust*) and Practice 3 (*commitment*). Strong alignment was also found among OMB SPD 2a and NASEM Principle 4 (*independence*) and Practice 10 (*transparency*).

*Areas of Tension:* We also observe areas of tension. Although the elements of each guidance generally align, we observed consistent tension among NASEM Practice 7 (*collaboration*) across OMB SPD 1, 2, and 2a. We also observed consistent tension: OMB SPD 1 Responsibility 1 (*relevance*) with five of ten NASEM practices (*mission, authority, collaboration, respect, transparency*).[26] To a lesser extent, this tension also occurs between OMB SPD 1 Responsibility 4 (*trust*) with three of NASEM practices (*research, collaboration,*

---

[25] Key: Areas of alignment are indicated by ✓ (check). Areas of conflict are indicated by ~ (not). Areas of tension (both alignment and potential conflict) are indicated by ✓ (check) and ~ (not). Patterns of alignment are shaded in green. Patterns of tension are shaded in yellow. Patterns of gaps in guidance are shaded in gray.

[26] These tensions are observed commonly in federal statistical practice. For example, tension may arise when practitioners are asked for timely or public interest data (Responsibility 1) if the activity is not explicitly allowable given the agency's mission (Practice 1).

*users' needs*).[27]

| OMB SPD Guidelines | PRI1 Relevance | PRI2 Credibility | PRI3 Trust | PRI4 Independence | PRI5: Improvement | PRA1: Mission | PRA2: Authority | PRA3: Commitment | PRA4: Staff | PRA5: Research | PRA6: Evaluation | PRA7: Collaboration | PRA8: Respect | PRA9: Users' Needs | PRA10: Transparency |
|---|---|---|---|---|---|---|---|---|---|---|---|---|---|---|---|
| 1.1 Relevant | ✓ | ✓ | ✓ | | | (~) | (~) | | (✓) | | (~) | (~) | ✓ | (~) |
| 1.2 Credible and accurate | | ✓ | ✓ | ✓ | ✓ | ✓ | ✓ | ✓ | ✓ | ✓ | (~) | (✓) | ✓(~) | ✓ |
| 1.3 Objective | | ✓ | ✓ | ✓ | ✓ | | | ✓ | ✓ | ✓ | (~) | (✓) | ✓(~) | ✓ |
| 1.4 Trust | ✓ | ✓ | ✓ | | | ✓ | ✓ | ✓ | (~) | | (~) | ✓ | (~) | |
| 2.1.1 Written plan | | ✓ | ✓ | | (✓) | ✓ | | | ✓ | | | ✓~ | | ✓ | |
| 2.1.2 Survey design | | ✓ | ✓ | | (✓) | | | | ✓ | | | ✓~ | | ✓ | ✓ |
| 2.1.3 Response rate design | | ✓ | ✓ | | (✓) | ✓ | | | ✓ | | ✓ | ✓~ | (✓) | ✓ | ✓ |
| 2.1.4 Functioning components | (✓) | ✓ | ✓ | | ✓ | ✓ | ✓ | | ✓ | | ✓ | ✓~ | ✓ | ✓ | ✓ |
| 2.2.1 Appropriate frame | (✓) | ✓ | ✓ | | | ✓ | | | ✓ | | ✓ | ✓~ | | | ✓ |
| 2.2.2 Notify respondents | | ✓ | ✓ | | | (✓) | ✓ | ✓ | | | ✓ | | | ✓ | ✓ |
| 2.2.3 Balance quality v burden | | ✓ | ✓ | (✓)(~) | ✓ | ✓ | | | ✓ | | ✓ | ✓~ | ✓ | | ✓ |
| 2.3.1 Appropriate data edits | (~) | ✓ | ✓ | (✓)(~) | | ✓ | (✓) | | ✓ | | ✓ | ✓~ | (✓) | | ✓ |
| 2.3.2 Nonresponse analysis | (✓) | ✓ | ✓ | | (✓) | ✓ | | | ✓ | | | | | (✓) | ✓ |
| 2.3.3 Quality for other study | (~) | ✓ | ✓ | | | (✓)(~) | | | ✓ | | ✓ | ✓ | ~ | ✓ | ✓ |
| 2.3.4 Avoid disclosure | | ✓ | ✓ | | | (✓) | | | ✓ | | ✓ | ✓~ | ~ | ~ | ✓ |
| 2.3.5 Evaluate data quality | (✓) | ✓ | ✓ | | | | | | ✓ | | ✓ | ✓~ | ✓ | ✓ | ✓ |
| 2.4.1 Use theory and methods | | ✓ | ✓ | | (✓) | | ✓ | ✓ | | | | ✓~ | | ✓ | ✓ |
| 2.5.1 Analysis plan | (✓) | ✓ | ✓ | (✓)(~) | (✓) | | ✓ | ✓ | | | | ✓~ | | ✓ | ✓ |
| 2.5.2 Good statistical practice | | ✓ | ✓ | | (✓)(~) | ✓ | | | ✓ | ✓~ | ✓ | ✓~ | ✓ | ✓ | |
| 2.6.1 Review dissemination | ✓ | ✓ | ✓ | (✓)(~) | ✓ | ✓ | (✓) | | ✓ | | ✓ | ✓~ | | ✓ | ✓ |
| 2.7.1 Equitable dissemination | (✓)(~) | ✓ | ✓ | (✓) | ✓ | | (✓) | | ✓ | | ✓ | ✓~ | ✓ | ✓ | ✓ |
| 2.7.2 Data protection | | ✓ | ✓ | | (✓)(~) | | | | ✓ | | ✓ | ✓~ | ✓ | ✓ | ✓~ |
| 2.7.3 Survey documentation | | ✓ | ✓ | (✓) | ✓ | | | | ✓ | | ✓ | ✓ | | ✓ | ✓ |
| 2.7.4 Public use microdata | (✓) | ✓ | ✓ | | ✓ | | | | ✓ | | (✓) | ✓~ | ✓~ | ✓ | ✓ |
| 2.A.1 Methodological plan | | ✓ | ✓ | ✓ | | ✓ | | | ✓ | | (✓) | ~ | | ✓ | ✓ |
| 2.A.2 Sample selection standard | | ✓ | ✓ | (✓) | | | | | ✓ | | (✓) | (~) | ✓ | | ✓ |
| 2.A.3 Interview guide standard | | ✓ | ✓ | ✓ | | ✓ | | | ✓ | | (✓) | (~) | (✓) | | ✓ |
| 2.A.4 Systematic analysis | | ✓ | ✓ | ✓ | (✓) | | (✓) | | ✓ | | | (~) | | | ✓ |
| 2.A.5 Transparent analysis | | ✓ | ✓ | ✓ | | | ✓ | | ✓ | | ✓ | ✓ | ✓ | ✓ | ✓ |
| 2.A.6 Final reports standard | | ✓ | (✓) | ✓ | | | | | ✓ | | ✓ | (~) | ✓ | ✓ | ✓ |
| 2.A.7 Reporting results standard | | ✓ | (✓) | ✓ | | | | | ✓ | | ✓ | (~) | | ✓ | ✓ |

*Table a Summary of themes identified in OMB SPD 1, 2, 2a and NASEM PNP*

*Guidance Gaps*: A consistent guidance gap was observed across OMB SPD 1, 2 and 2a with regard to NASEM Practices 4 and 5 (*advancement of staff*, *an active research program*), shaded in gray. Without explicit recognition across guidance documents, federal agencies may have greater difficulty supporting staff development and research–both key aspects of continuous improvement to support relevant and credible statistics.

*Discussion:* Overall, our method shows general alignment between SPD 1, 2, 2a and NASEM PNP. Alignment is particularly strong in guidance areas: *credibility, objectivity, trust,*

---

[27] These tensions are also familiar in federal statistical practice. For example, supporting robust research programs (Practice 5) and users' needs (Practice 9) should not be prioritized over the rights of data providers.

*independence, and transparency*. These areas of strong alignment across guidance documents provide a particularly solid basis for strengthening existing guidance through higher level authority, such as regulation or law.

However, given that the target audiences for these guidance documents are similar, we did not observe the breadth and strength of alignment we anticipated. We observed consistent areas of tension across guidance documents, particularly within the intersection of OMB SPD Responsibilities 1 (relevance) and 4 (trust) and several NASEM practices. Additionally, we observed a consistent guidance gap in the *advancement of staff* as well as *an active research program* across SPD 1, 2, and 2a. These tensions and gap could be addressed through clarification of language when implementing routine reviews and possible revision of these guidance documents. As discussed above, because the ASA EG describe professional expectations for how practices should be implemented (or behaviors), they may be particularly useful to such reviews. Areas of tension should be resolved prior to elevating guidance to higher levels of authority, such as regulation or law.

b. OMB Data Ethics Tenets and NASEM Principles and Practices

| NASEM Principles and Practices (2021): / OMB Data Ethics Tenets (2020): | T1 Uphold laws and regulations | T2 Respect public | T3 Respect privacy and confidentiality | T4 Integrity | T5 Accountability | T6 Transparency | T7 Stay informed |
|---|---|---|---|---|---|---|---|
| PRI1 Relevance | (~) | ✓ ~ | (~) | (~) | (~) | (~) | |
| PRI2 Credibility | ✓ | ✓ | ✓ | ✓ | ✓ | ✓ | ✓ |
| PRI3 Trust | ✓ | ✓ | ✓ | ✓ | ✓ | ✓ | ✓ |
| PRI4 Independence | ✓ | ✓ | ✓ | ✓ | ✓ | ✓ | |
| PRI5: Improvement | (✓) | | ✓ | | (✓) | | ✓ |
| PRA1: Mission | (✓) | | | | | ✓ | |
| PRA2: Authority | ✓ | | (✓) | ✓ | ✓ | ✓ | |
| PRA3: Commitment | ✓ | ✓ | ✓ | ✓ | ✓ | ✓ | (✓) |
| PRA4: Staff | ✓ | | | ✓ | ✓ | | ✓ |
| PRA5: Research | (✓) | | (~) | | | (✓) | ✓ |
| PRA6: Evaluation | ✓ | ✓ | | | ✓ | ✓ | |
| PRA7: Collaboration | ~ | ~ | (~) | (~) | (~) | (~) | ✓ |
| PRA8: Respect | ✓ | ✓ | ✓ | | ✓ | ✓ | (✓) |
| PRA9: Users' Needs | ~ | ~ | (~) | (~) | (~) | ✓ ~ | (✓) |
| PRA10: Transparency | ✓ | ✓ | ✓ | ✓ | ✓ | ✓ | (✓) |

*Table b Summary of themes identified in OMB Data Ethics Tenets and NASEM Principles and Practices*

*Target Audience:* Recall that the target audiences for OMB DET and NASEM PNP differ. OMB DET are intended to communicate broad expectations of leadership in federal agencies engaged in the data lifecycle. NASEM PNP are intended to communicate broad expectations in leadership of federal statistical agencies–a subset within the target audience of OMB DET.

*Areas of Strong Alignment:* See Table b. [28]We find that NASEM Principles 2 (*credibility*) and 3 (*trust*) are fully aligned with all OMB DET. Additionally, NASEM Practices 3 (*commitment*) and 10 (*transparency*) are fully aligned with all OMB DET. Alignment with NASEM Principle 4 (*independence*) is also notable.

*Areas of Tension:* We observe NASEM Principle 1 (*relevance*) Practice 7 (*collaboration*), and Practice 9 (*users' needs*) must be accomplished in balance with OMB DET 1-6 to avoid data collection or dissemination of statistical products that have the potential to harm (or perception of harm) to segments of society.

---

[28] Key: Areas of alignment are indicated by ✓ (check). Areas of conflict are indicated by ~ (not). Areas of tension (both alignment and potential conflict) are indicated by ✓ (check) and ~ (not). Patterns of alignment are shaded in green. Patterns of tension are shaded in yellow. Patterns of gaps in guidance are shaded in gray.

*Guidance Gap*: None observed.

*Discussion:* Overall, our method shows general and, in some areas, strong alignment between NASEM PNP and OMB DET. Alignment is particularly strong in guidance areas: *credibility, trust, commitment, transparency, and independence*. These areas of strong alignment across guidance documents provide a particularly solid basis for strengthening existing guidance through higher level authority, such as regulation or law.

Even though the target audiences for these guidance documents differ somewhat, tensions were observed in only particular areas: *relevance, collaboration*, and *users' needs*. We did not observe guidance gaps. Tensions could be addressed through clarification of language when implementing routine reviews and possible revision of these guidance documents. As discussed above, because the ASA EG describe professional expectations for how practices should be implemented (or behaviors), they may be particularly useful to such reviews. Areas of tension should be resolved prior to elevating guidance to higher levels of authority, such as regulation or law.

c. **OMB Data Ethics Tenets and OMB Statistical Policy Directives 1, 2 and 2a**

| OMB SPD Guidelines | T1 Uphold laws and regulations | T2 Respect public | T3 Respect privacy and confidentiality | T4 Integrity | T5 Accountability | T6 Transparency | T7 Stay informed |
|---|---|---|---|---|---|---|---|
| 1.1 Relevant | (~) | (~) | (~) | (~) | (~) | ✓ | (✓) |
| 1.2 Credible and accurate | ✓ | ✓ | ✓ | ✓ | ✓ | ✓ | (✓) |
| 1.3 Objective | | | | | | ✓ | |
| 1.4 Trust | ✓ | ✓ | ✓ | ✓ | ✓ | ✓ | (✓) |
| 2.1.1 Written plan | ✓ | | | ✓ | ✓ | ✓ | (✓) |
| 2.1.2 Survey design | ✓ | | | ✓ | ✓ | | (✓) |
| 2.1.3 Response rate design | ✓ | ✓ | | ✓ | ✓ | ✓ | (✓) |
| 2.1.4 Functioning components | ✓ | ✓ | (✓) | ✓ | ✓ | ✓ | (✓) |
| 2.2.1 Appropriate frame | ✓ | (✓) | | ✓ | ✓ | | |
| 2.2.2 Notify respondents | ✓ | ✓ | ✓ | ✓ | ✓ | | |
| 2.2.3 Balance quality v burden | ✓ | (✓) | | ✓ | ✓ | | (✓) |
| 2.3.1 Appropriate data edits | ✓ | (✓) | | ✓ | ✓ | ✓ | (✓) |
| 2.3.2 Nonresponse analysis | ✓ | | | ✓ | ✓ | ✓ | (✓) |
| 2.3.3 Quality for other study | ✓ | ✓ | ✓ | ✓ | ✓ | ✓ | |
| 2.3.4 Avoid disclosure | ✓ | ✓ | ✓ | ✓ | ✓ | ✓ | |
| 2.3.5 Evaluate data quality | ✓ | (✓) | | ✓ | ✓ | ✓ | |
| 2.4.1 Use theory and methods | ✓ | | | ✓ | ✓ | ✓ | ✓ |
| 2.5.1 Analysis plan | ✓ | | | | ✓ | ✓ | |
| 2.5.2 Good statistical practice | ✓ | | (✓) | | ✓ | ✓ | ✓ |
| 2.6.1 Review dissemination | ✓ | (~) | ✓ | ✓ | ✓ | | |
| 2.7.1 Equitable dissemination | ✓ | ✓ | | ✓ | ✓ | ✓ | (✓) |
| 2.7.2 Data protection | ✓ | ✓ | ✓ | ✓ | ✓ | ✓ | |
| 2.7.3 Survey documentation | ✓ | | | | | ✓ | |
| 2.7.4 Public use microdata | ✓ | ✓ | ✓ | | | ✓ | |
| 2.A.1 Methodological plan | ✓ | | | ✓ | ✓ | ✓ | (✓) |
| 2.A.2 Sample selection standard | ✓ | (✓) | | ✓ | ✓ | ✓ | |
| 2.A.3 Interview guide standard | (✓) | | | | ✓ | ✓ | |
| 2.A.4 Systematic analysis | ✓ | | | | | | (✓) |
| 2.A.5 Transparent analysis | ✓ | ✓ | ✓ | ✓ | ✓ | ✓ | |
| 2.A.6 Final reports standard | ✓ | | | | ✓ | ✓ | |
| 2.A.7 Reporting results standard | ✓ | (~) | (~) | | ✓ | ✓ | |

*Table c Summary of themes identified in OMB Statistical Policy Directives 1, 2, 2a and OMB Data Ethics Tenets*

*Target Audience:* Recall that the target audiences for OMB DET and OMB SPD differ. OMB DET are intended to communicate broad expectations of leadership in federal agencies engaged in the data lifecycle. OMB SPD are intended to communicate broad expectations in leadership of federal statistical agencies–a subset of the target audience of OMB DET.

*Areas of Strong Alignment:* See Table c.[29] We observe strong and general alignment across OMB DET and SPD 1, 2, 2a. Particularly strong is alignment with OMB DET Tenet 1 (uphold law) with all SPD elements. Also notable are alignment with OMB DET Tenets 4, 5, 6, and 7 (*integrity, accountability, transparency, and informed*) across all SPD elements.

---

[29] Key: Areas of alignment are indicated by ✓ (check). Areas of conflict are indicated by ~ (not). Areas of tension (both alignment and potential conflict) are indicated by ✓ (check) and ~ (not). Patterns of alignment are shaded in green. Patterns of tension are shaded in yellow. Patterns of gaps in guidance are shaded in gray.

Alignment of several elements of SPD 2 with (almost) all OMB DET is also notable: *functioning* components, *notify respondents*, *quality*, *avoid disclosure*, *equitable dissemination*, and *data protection*.

*Areas of Tension*: We observe tensions between OMB SPD Responsibility 1 (*relevant*) and OMB DET 1-5. In these cases, the practitioner must prioritize *upholding law*, *respect for the public*, *respect for privacy*, *integrity,* and *accountability* when these are in conflict with *relevance (timeliness)*.

*Guidance Gaps*: Two guidance gaps in particular are observed with our method. OMB DET could benefit from additional language regarding objectivity of data. OMB SPD 2, and 2a could benefit from additional language regarding privacy and confidentiality of data.

*Discussion:* We observe strong and general alignment across OMB DET and SPD 1, 2, 2a. Particularly strong is alignment with *uphold law*, *integrity, accountability, transparency, and informed*. With regard to SPD 2, we also note strong alignment with *functioning* components, *notify respondents*, *quality*, *avoid disclosure*, *equitable dissemination*, and *data protection*. These areas of strong alignment across guidance documents provide a particularly solid basis for strengthening existing guidance through higher level authority, such as regulation or law.

Even though the target audiences for these guidance documents differ somewhat, tensions were limited to the area of *relevance*. In these cases, the practitioner must prioritize *upholding law*, *respect for the public*, *respect for privacy*, *integrity,* and *accountability* when these are in conflict with *relevance (timeliness)*. Two guidance gaps are observed. Tensions and gaps could be addressed through clarification of language when implementing routine reviews and possible revision of these guidance documents. As discussed above, because the ASA EG describe professional expectations for how practices should be implemented (or behaviors), they may be particularly useful to such reviews. Areas of tension should be resolved prior to elevating guidance to higher levels of authority, such as regulation or law.

d. Overall alignment

Table 4 presents the overall summary of alignment of the ASA Ethical Guidelines and the federal guidelines examined here. Overall, the ASA EG align well with federal guidance. Although there is not complete alignment of federal guidance and ethical practice, every ASA EG principle is aligned with at least one federal guideline (usually two). Notably, the ASA EG

| ASA Ethical Guidelines / Comparison Guidelines | A Accountability | B Integrity | C Stakeholder | D Subjects | E Other Disciplines | F Other Statisticians | G Leadership | H Misconduct |
|---|---|---|---|---|---|---|---|---|
| OMB Statistical Policy Directive 1 | Tension | Alignment | Alignment | Tension | Alignment | Alignment | Tension | Alignment |
| NASEM Principles and Practices | Tension | Tension | Tension | Tension | Tension | Tension | Tension | Tension |
| OMB Data Ethics Tenets | Alignment | Alignment | Alignment | Alignment | Alignment | Alignment | Alignment | Alignment |

*Table 4 Overall alignment of ASA Ethical Guidelines with examined federal guidelines*

are aligned consistently across OMB DET.

Looking more closely at the conceptual level, we observe sources of tension across guidelines. Across all guidances examined, there is clearest alignment for credible, objective, transparent statistics provided in the public's trust. These expectations are clearly defined, widely-held, and thus might be considered core values uniting national statistics and data science professions.

However, we also observe common areas of tension across guidelines. To resolve these tensions, the guidance areas of relevance, respect for data providers, collaboration, and addressing users' needs may require prioritization (that is, giving some guidelines greater weight than others, such as the core values noted above) and greater clarity (particularly at the activity and conduct level). Moreover, the complexities arising from the alignment and identification of potential tensions across federal guidance documents themselves should be considered in training and onboarding of new practitioners. The ASA EG may be particularly valuable in resolving these tensions by providing examples at the activity and conduct level.

### E. Discussion

In this analysis, we explored the extent to which the ASA EG support accomplishment of the principles, practices, and tenets outlined in current federal guidance for official statistics in the United States. We examined four key sources of national guidance in the professional conduct of statistics and data science. Each of these guidances was developed for somewhat different purposes, and therefore, somewhat different audiences. It is therefore reasonable to anticipate differences in alignment.

Despite these differences of audience and purpose, we found broad alignment of guidelines overall. Even taking into account the time spanned across releases, these guidelines signal a common understanding of ethical statistical practice and its importance. At the element level, areas of alignment were strongest pertaining to credibility, transparency, and trust. Each guideline examined included these elements. These are clearly areas where federal guidance is well aligned with ethical practice standards. These areas would have the most solid basis for strengthening existing guidance by elevating authority, such as by law or regulation. Elements of independence, relevance, accountability, and improvement were also common, but to a lesser degree. The mandate for these areas may be established in law, such as the Evidence Act, but further refinement and clarification may be needed when developing regulation.

Across federal guidance, we found the ASA Ethical Guidelines align most closely with the OMB DET. We also found the ASA EG strongly align with both OMB SPD 1, 2, and 2a and NASEM PNP. Nonetheless, many potential tensions were identified. OMB SPD 1 and NASEM PNP were strongly aligned, but there were more tensions observed between SPD 2 and 2a and NASEM practices than may have been anticipated, given the jointly shared target audience of these guidances. Both OMB SPD 1, 2, and 2a and NASEM PNP aligned with the OMB DET, but to a lesser extent than the ASA EG.

A closer examination identified sources of potential tension both within and across these federal guidances. The majority of these tensions arise in pursuing relevant and timely work while adhering to other guidance elements, such as credibility, trust, and independence. There are many ways in which these elements may be undermined if practitioners prioritize timeliness, or sharing resources. These tensions, while not explicit within a given set of federal guidelines, are undoubtedly familiar to federal statisticians and data science practitioners. Nonetheless, it can be beneficial to place these potential tensions in context. These are two situations where nonpractitioners and practitioners are most likely to intersect, and where expectations may differ. In the extreme case, there may be demands to prioritize expediency over accuracy, or relevance over privacy. Examination of the ASA EG can assist in resolving these tensions by providing guidance at the level of activity and conduct. Periodic reviews of federal guidances are opportunities to clarify existing language. Apparent tensions should be resolved in federal guidance language before elevating authority, such as law or regulation.

We underscore that apparent tensions provide opportunities for greater dialogue across disciplines. Consider in particular NASEM Practice 7 (collaboration). Increasingly, and in particular since the authorization of the Evidence Act, coordination of the US federal statistical system has broadened its scope to more formally include other federal agencies with significant statistical and data science programs, but whose core mission is not focused solely on production of statistics. As this examination of guidelines itself signals, coordination across statistical agencies and other agencies has the potential to create tension for practitioners, in the sense that agencies pursuing coordination (even to achieve common goals) might differ in priorities assigned to (even shared) principles and their approach to achieving them (practices). Accordingly, we see a role for a broader discussion of the role of professional ethics, such as the ASA EG in resolving potential tensions across federal agencies with statistical and data science programs.

OMB SPD, OMB DET, and (to a lesser extent) NASEM PNP are aspirational by design. They are intended to signal expectations of federal statistics and data science professionals to the public at large. In this way, they are well suited to inform regulation and policy. In practice, however, aspirations are very difficult to fully meet. Balance among (potentially) competing aspirations is needed. Indeed, a "one size fits all" approach to ethical statistical practice is not the aim of these guidelines; rather, they are intended to provide frameworks for critical decision-making that must be context-specific.

It is in this space that the ASA Ethical Guidelines are an especially powerful resource. Because they are focused on action and conduct (by design), compliance is easier to determine, and encouragement is easier to implement. Accordingly, avoiding (or resolving) potential sources of tension may be more successful where expectations and priorities are identified first at the goal level (where broader alignment is found across guidelines), and (perhaps) ultimately addressed through specific practices and behaviors. The ASA EG describe many diverse ways to *support* credibility, trust, and independence–and also describe many ways that these can be undermined.

### F. Conclusion

We found the alignment model useful to identify areas of commonality across guidelines developed for related but distinct audiences who engage in statistical and data science practices in the U.S. federal setting. The ASA EG are supportive of the federal guidelines examined here. This demonstrates strong shared professional goals. Further, the alignment model suggests that where goals align, goals articulated by a particular guideline (such as OMB SPD 1) can be promoted by the practices described in another guideline (such as ASA EG). The ASA EG are particularly useful because they specify behaviors that can support quality in practices and principles across all national guidelines examined here.

However, areas of potential tension within and across guidelines were also found. These tensions, while perhaps not explicit in the guidelines themselves, are well-known to the statistical community. It is therefore clear that no single federal guideline document, on its own, would be sufficient to inform future statistical or data science policy. Indeed, we found apparent conflicts and guidance gaps across federal guidances examined here. We note that OMB DET are most strongly and consistently supported by the ASA EG, and strongly align with OMB SPD and NASEM PNP. However, federal guidance, due to its brevity for communication purposes, does not provide guidance at the activity and behavior level that would be necessary to resolve apparent tensions. Instead, it would be most instructive to use the tensions identified here to guide a balanced approach toward often-competing goals. The alignment model helps identify the specific goals, practices, and behaviors that must be considered in developing this balance, and the ASA EG can serve as an ethical practice standard to support this balance.

Tensions signal where greater attention should be paid to challenging (implicit) assumptions, including priorities among shared goals. Tensions also signal where greater dialogue across disciplines and audiences may be warranted, and where examples (and counter examples) may be most useful in supporting ethical practice - and for training new members of the community to promote ethical practice and decision making. By signaling the need for this balance, this study of alignment between these largely complementary guidelines can foster greater dialogue within the community of statistics practitioners. Ultimately, resolving these tensions will be context driven. The ASA EG are well-positioned to resolve these potential tensions. The examples of practices and behaviors to avoid and pursue can be instructive not only to students of statistics and data science, but also to new employees as well as established practitioners crafting new policies for ethical use of data and statistical and data science practices in evidence building.

G.  Future work

In a separate paper (Tractenberg & Park, 2023), we use the same alignment method and model to examine congruence in guidelines from the ASA and diverse international organizational bodies (ASA Ethical Guidelines and the UN Fundamental Principles of Official Statistics, OECD Good Statistical Practice, and European Statistics Code of Practice). We also consider comparison of ASA EG to the principles identified in OMB M-20-12.

Annex
1 ASA Ethical Guidelines
2 Detail Tables 1-3; a-g

Annex 1. ASA Ethical Guidelines for Statistical Practice, 2022

**Ethical Guidelines for Statistical Practice**
*Prepared by the Committee on Professional Ethics
of the American Statistical Association*
**February 2022**

**PURPOSE OF THE GUIDELINES:**

The American Statistical Association's Ethical Guidelines for Statistical Practice are intended to help statistical practitioners make decisions ethically. In these Guidelines, "statistical practice" includes activities such as: designing the collection of, summarizing, processing, analyzing, interpreting, or presenting, data; as well as model or algorithm development and deployment. Throughout these Guidelines, the term "statistical practitioner" includes all those who engage in statistical practice, regardless of job title, profession, level, or field of degree. The Guidelines are intended for individuals, but these principles are also relevant to organizations that engage in statistical practice.

The Ethical Guidelines aim to promote accountability by informing those who rely on any aspects of statistical practice of the standards that they should expect. Society benefits from informed judgments supported by ethical statistical practice. All statistical practitioners are expected to follow these Guidelines and to encourage others to do the same.

In some situations, Guideline principles may require balancing of competing interests. If an unexpected ethical challenge arises, the ethical practitioner seeks guidance, not exceptions, in the Guidelines. To justify unethical behaviors, or to exploit gaps in the Guidelines, is unprofessional, and inconsistent with these Guidelines.

### PRINCIPLE A: Professional Integrity and Accountability

Professional integrity and accountability require taking responsibility for one's work. Ethical statistical practice supports valid and prudent decision making with appropriate methodology. The ethical statistical practitioner represents their capabilities and activities honestly, and treats others with respect.

**The ethical statistical practitioner:**

1. Takes responsibility for evaluating potential tasks, assessing whether they have (or can attain) sufficient competence to execute each task, and that the work and timeline are feasible. Does not solicit or deliver work for which they are not qualified, or that they would not be willing to have peer reviewed.
2. Uses methodology and data that are valid, relevant, and appropriate, without favoritism or prejudice, and in a manner intended to produce valid, interpretable, and reproducible results.
3. Does not knowingly conduct statistical practices that exploit vulnerable populations or create or perpetuate unfair outcomes.
4. Opposes efforts to predetermine or influence the results of statistical practices, and resists pressure to selectively interpret data.
5. Accepts full responsibility for their own work; does not take credit for the work of others; and gives credit to those who contribute. Respects and acknowledges the intellectual property of others.
6. Strives to follow, and encourages all collaborators to follow, an established protocol for authorship. Advocates for recognition commensurate with each person's contribution to the work. Recognizes that inclusion as an author does imply, while acknowledgement may imply, endorsement of the work.
7. Discloses tensions of interest, financial and otherwise, and manages or resolves them according to established policies, regulations, and laws.
8. Promotes the dignity and fair treatment of all people. Neither engages in nor condones discrimination based on personal characteristics. Respects personal boundaries in interactions and avoids harassment including sexual harassment, bullying, and other abuses of power or authority.
9. Takes appropriate action when aware of deviations from these Guidelines by others.
10. Acquires and maintains competence through upgrading of skills as needed to maintain a high standard of practice.
11. Follows applicable policies, regulations, and laws relating to their professional work, unless there is a compelling ethical justification to do otherwise.
12. Upholds, respects, and promotes these Guidelines. Those who teach, train, or mentor in statistical practice have a special obligation to promote behavior that is consistent with these Guidelines.

### PRINCIPLE B: Integrity of Data and Methods

The ethical statistical practitioner seeks to understand and mitigate known or suspected limitations, defects, or biases in the data or methods and communicates potential impacts on the interpretation, conclusions, recommendations, decisions, or other results of statistical practices.
**The ethical statistical practitioner:**

1.  Communicates data sources and fitness for use, including data generation and collection processes and known biases. Discloses and manages any tensions of interest relating to the data sources. Communicates data processing and transformation procedures, including missing data handling.

2.  Is transparent about assumptions made in the execution and interpretation of statistical practices including methods used, limitations, possible sources of error, and algorithmic biases. Conveys results or applications of statistical practices in ways that are honest and meaningful.

3.  Communicates the stated purpose and the intended use of statistical practices. Is transparent regarding a priori versus post hoc objectives and planned versus unplanned statistical practices. Discloses when multiple comparisons are conducted, and any relevant adjustments.

4.  Meets obligations to share the data used in the statistical practices, for example, for peer review and replication, as allowable. Respects expectations of data contributors when using or sharing data. Exercises due caution to protect proprietary and confidential data, including all data that might inappropriately harm data subjects.

5.  Strives to promptly correct substantive errors discovered after publication or implementation. As appropriate, disseminates the correction publicly and/or to others relying on the results.

6.  For models and algorithms designed to inform or implement decisions repeatedly, develops and/or implements plans to validate assumptions and assess performance over time, as needed. Considers criteria and mitigation plans for model or algorithm failure and retirement.

7.  Explores and describes the effect of variation in human characteristics and groups on statistical practice when feasible and relevant.

## PRINCIPLE C: Responsibilities to Stakeholders

Those who fund, contribute to, use, or are affected by statistical practices are considered stakeholders. The ethical statistical practitioner respects the interests of stakeholders while practicing in compliance with these Guidelines.

**The ethical statistical practitioner:**
1. Seeks to establish what stakeholders hope to obtain from any specific project. Strives to obtain sufficient subject-matter knowledge to conduct meaningful and relevant statistical practice.
2. Regardless of personal or institutional interests or external pressures, does not use statistical practices to mislead any stakeholder.
3. Uses practices appropriate to exploratory and confirmatory phases of a project, differentiating findings from each so the stakeholders can understand and apply the results.
4. Informs stakeholders of the potential limitations on use and re-use of statistical practices in different contexts and offers guidance and alternatives, where appropriate, about scope, cost, and precision considerations that affect the utility of the statistical practice.
5. Explains any expected adverse consequences from failing to follow through on an agreed-upon sampling or analytic plan.
6. Strives to make new methodological knowledge widely available to provide benefits to society at large. Presents relevant findings, when possible, to advance public knowledge.
7. Understands and conforms to confidentiality requirements for data collection, release, and dissemination and any restrictions on its use established by the data provider (to the extent legally required). Protects the use and disclosure of data accordingly. Safeguards privileged information of the employer, client, or funder.
8. Prioritizes both scientific integrity and the principles outlined in these Guidelines when interests are in conflict.

## PRINCIPLE D: Responsibilities to Research Subjects, Data Subjects, or those directly affected by statistical practices

The ethical statistical practitioner does not misuse or condone the misuse of data. They protect and respect the rights and interests of human and animal subjects. These responsibilities extend to those who will be directly affected by statistical practices.

**The ethical statistical practitioner:**

1. Keeps informed about and adheres to applicable rules, approvals, and guidelines for the protection and welfare of human and animal subjects. Knows when work requires ethical review and oversight.[1]
2. Makes informed recommendations for sample size and statistical practice methodology in order to avoid the use of excessive or inadequate numbers of subjects and excessive risk to subjects
3. For animal studies, seeks to leverage statistical practice to reduce the number of animals used, refine experiments to increase the humane treatment of animals, and replace animal use where possible.
4. Protects people's privacy and the confidentiality of data concerning them, whether obtained from the individuals directly, other persons, or existing records. Knows and adheres to applicable rules, consents, and guidelines to protect private information.

5. Uses data only as permitted by data subjects' consent when applicable or considering their interests and welfare when consent is not required. This includes primary and secondary uses, use of repurposed data, sharing data, and linking data with additional data sets.
6. Considers the impact of statistical practice on society, groups, and individuals. Recognizes that statistical practice could adversely affect groups or the public perception of groups, including marginalized groups. Considers approaches to minimize negative impacts in applications or in framing results in reporting.
7. Refrains from collecting or using more data than is necessary. Uses confidential information only when permitted and only to the extent necessary. Seeks to minimize the risk of re-identification when sharing de-identified data or results where there is an expectation of confidentiality. Explains any impact of de-identification on accuracy of results.
8. To maximize contributions of data subjects, considers how best to use available data sources for exploration, training, testing, validation, or replication as needed for the application. The ethical statistical practitioner appropriately discloses how the data is used for these purposes and any limitations.
9. Knows the legal limitations on privacy and confidentiality assurances and does not over-promise or assume legal privacy and confidentiality protections where they may not apply.
10. Understands the provenance of the data, including origins, revisions, and any restrictions on usage, and fitness for use prior to conducting statistical practices.
11. Does not conduct statistical practice that could reasonably be interpreted by subjects as sanctioning a violation of their rights. Seeks to use statistical practices to promote the just and impartial treatment of all individuals.

## PRINCIPLE E: Responsibilities to members of multidisciplinary teams

Statistical practice is often conducted in teams made up of professionals with different professional standards. The statistical practitioner must know how to work ethically in this environment.

**The ethical statistical practitioner:**

1. Recognizes and respects that other professions may have different ethical standards and obligations. Dissonance in ethics may still arise even if all members feel that they are working towards the same goal. It is essential to have a respectful exchange of views.
2. Prioritizes these Guidelines for the conduct of statistical practice in cases where ethical guidelines conflict.
3. Ensures that all communications regarding statistical practices are consistent with these Guidelines. Promotes transparency in all statistical practices.
4. Avoids compromising validity for expediency. Regardless of pressure on or within the team, does not use inappropriate statistical practices.

## PRINCIPLE F: Responsibilities to Fellow Statistical Practitioners and the Profession

Statistical practices occur in a wide range of contexts. Irrespective of job title and training, those who practice statistics have a responsibility to treat statistical practitioners, and the profession, with respect. Responsibilities to other practitioners and the profession include honest communication and engagement that can strengthen the work of others and the profession.

**The ethical statistical practitioner:**

1. Recognizes that statistical practitioners may have different expertise and experiences, which may lead to divergent judgments about statistical practices and results. Constructive discourse with mutual respect focuses on scientific principles and methodology and not personal attributes.
2. Helps strengthen, and does not undermine, the work of others through appropriate peer review or consultation. Provides feedback or advice that is impartial, constructive, and objective.
3. Takes full responsibility for their contributions as instructors, mentors, and supervisors of statistical practice by ensuring their best teaching and advising -- regardless of an academic or non-academic setting -- to ensure that developing practitioners are guided effectively as they learn and grow in their careers.
4. Promotes reproducibility and replication, whether results are "significant" or not, by sharing data, methods, and documentation to the extent possible.
5. Serves as an ambassador for statistical practice by promoting thoughtful choices about data acquisition, analytic procedures, and data structures among non-practitioners and students. Instills appreciation for the concepts and methods of statistical practice.

### PRINCIPLE G: Responsibilities of Leaders, Supervisors, and Mentors in Statistical Practice

Statistical practitioners leading, supervising, and/or mentoring people in statistical practice have specific obligations to follow and promote these Ethical Guidelines. Their support for – and insistence on – ethical statistical practice are essential for the integrity of the practice and profession of statistics as well as the practitioners themselves.

**Those leading, supervising, or mentoring statistical practitioners are expected to**:

1. Ensure appropriate statistical practice that is consistent with these Guidelines. Protect the statistical practitioners who comply with these Guidelines, and advocate for a culture that supports ethical statistical practice.
2. Promote a respectful, safe, and productive work environment. Encourage constructive engagement to improve statistical practice.
3. Identify and/or create opportunities for team members/mentees to develop professionally and maintain their proficiency.
4. Advocate for appropriate, timely, inclusion and participation of statistical practitioners as contributors/collaborators. Promote appropriate recognition of the contributions of statistical practitioners, including authorship if applicable.
5. Establish a culture that values validation of assumptions, and assessment of model/algorithm performance over time and across relevant subgroups, as needed. Communicate with relevant stakeholders regarding model or algorithm maintenance, failure, or actual or proposed modifications.

### PRINCIPLE H: Responsibilities Regarding Potential Misconduct

The ethical statistical practitioner understands that questions may arise concerning potential misconduct related to statistical, scientific, or professional practice. At times, a practitioner may accuse someone of misconduct, or be accused by others. At other times, a practitioner may be involved in the investigation of others' behavior. Allegations of misconduct may arise within different institutions with different standards and potentially different outcomes. The elements that follow relate specifically to allegations of statistical, scientific, and professional misconduct.

**The ethical statistical practitioner:**

1.  Knows the definitions of, and procedures relating to, misconduct in their institutional setting. Seeks to clarify facts and intent before alleging misconduct by others. Recognizes that differences of opinion and honest error do not constitute unethical behavior.
2.  Avoids condoning or appearing to condone statistical, scientific, or professional misconduct. Encourages other practitioners to avoid misconduct or the appearance of misconduct.
3.  Does not make allegations that are poorly founded, or intended to intimidate. Recognizes such allegations as potential ethics violations.
4.  Lodges complaints of misconduct discreetly and to the relevant institutional body. Does not act on allegations of misconduct without appropriate institutional referral, including those allegations originating from social media accounts or email listservs.
5.  Insists upon a transparent and fair process to adjudicate claims of misconduct. Maintains confidentiality when participating in an investigation. Discloses the investigation results honestly to appropriate parties and stakeholders once they are available.
6.  Refuses to publicly question or discredit the reputation of a person based on a specific accusation of misconduct while due process continues to unfold.
7.  Following an investigation of misconduct, supports the efforts of all parties involved to resume their careers in as normal a manner as possible, consistent with the outcome of the investigation.
8.  Avoids, and acts to discourage, retaliation against or damage to the employability of those who responsibly call attention to possible misconduct.

## APPENDIX
### Responsibilities of organizations/institutions

Whenever organizations and institutions design the collection of, summarize, process, analyze, interpret, or present, data; or develop and/or deploy models or algorithms, they have responsibilities to use statistical practice in ways that are consistent with these Guidelines, as well as promote ethical statistical practice.

**Organizations and institutions engage in, and promote, ethical statistical practice by**:

1.  Expecting and encouraging all employees and vendors who conduct statistical practice to adhere to these Guidelines. Promoting a workplace where the ethical practitioner may apply the Guidelines without being intimidated or coerced. Protecting statistical practitioners who comply with these Guidelines.
2.  Engaging competent personnel to conduct statistical practice, and promote a productive work environment.
3.  Promoting the professional development and maintenance of proficiency for employed statistical practitioners.
4.  Supporting statistical practice that is objective and transparent. Not allowing organizational objectives or expectations to encourage unethical statistical practice by its employees.

5. Recognizing that the inclusion of statistical practitioners as authors, or acknowledgement of their contributions to projects or publications, requires their explicit permission because it may imply endorsement of the work.
6. Avoiding statistical practices that exploit vulnerable populations or create or perpetuate discrimination or unjust outcomes. Considering both scientific validity and impact on societal and human well-being that results from the organization's statistical practice.
7. Using professional qualifications and contributions as the basis for decisions regarding statistical practitioners' hiring, firing, promotion, work assignments, publications and presentations, candidacy for offices and awards, funding or approval of research, and other professional matters.

**Those in leadership, supervisory, or managerial positions who oversee statistical practitioners promote ethical statistical practice by following Principle G and:**

8. Recognizing that it is contrary to these Guidelines to report or follow only those results that conform to expectations without explicitly acknowledging competing findings and the basis for choices regarding which results to report, use, and/or cite.
9. Recognizing that the results of valid statistical studies cannot be guaranteed to conform to the expectations or desires of those commissioning the study or employing/supervising the statistical practitioner(s).
10. Objectively, accurately, and efficiently communicating a team's or practitioners' statistical work throughout the organization.
11. In cases where ethical issues are raised, representing them fairly within the organization's leadership team.
12. Managing resources and organizational strategy to direct teams of statistical practitioners along the most productive lines in light of the ethical standards contained in these Guidelines.

---

[1] Examples of ethical review and oversight include an Institutional Review Board (IRB), an Institutional Animal Care and Use Committee (IACUC), or a compliance assessment.

# Annex 2. Detail Tables 1-3; a-g

Table 1. Alignment of ASA Ethical Guidelines with OMB Statistical Policy Directives 1, 2 and 2a

| ASA Ethical Guidelines (2022):<br><br>OMB Statistical Policies<br>Directive 1 (2014)<br>Directive 2 (2008)<br>Directive 2a (2016) | | A<br>Professional Integrity and Accountability | B<br>Integrity of Data and Methods | C<br>Stakeholders | D<br>Research Subjects/Data Subjects and Those Affected by Statistical Practices | E<br>Interdisciplinary Team Members | F<br>Other Practitioners/ Profession | G<br>Leader/ Supervisor/ Mentor and APPENDIX | H<br>Allegations of Potential Misconduct |
|---|---|---|---|---|---|---|---|---|---|
| SPD 1 | Responsibility 1: Produce and disseminate relevant and timely information | A2 ~A3, ~A4 (1) (2) | B1, B2, B3, B6 (4) | C1, C2, C8 (3) | D, D2, ~D10, ~D11 (D1) (D2) | E4 (1) | | G1, G5 (2) APPENDIX ~6, ~8, ~9 (3) | H2 (1) |
| | Responsibility 2: Conduct credible and accurate statistical activities | A, A2, A3, A4, A11, A12 (5) | B, B1, B2, B3 (3) | C, C1, C2, C4, C8 (4) | D1, D2, D4, D5, D6, D8, D10, D11 (8) | E2, E3, E4 (3) | F4 (1) | G1, G2, G5 (3) APPENDIX 1, 2, 4, 6, 8, 9, 11 (7) | H2 (1) |
| | Responsibility 3: Conduct objective statistical activities | A, A2, A3, A4, A11, A12 (5) | B, B1, B2, B3 (3) | C, C1, C2, C4, C8 (4) | D1, D2, D4, D5, D6, D8, D10, D11 (8) | E2, E3, E4 (3) | F4 (1) | G1, G2, G5 (3) APPENDIX 1, 2, 4, 6, 8, 9, 11 (7) | H2 (1) |
| | Responsibility 4: Protect the trust of information providers by ensuring the confidentiality and exclusive statistical use of their responses | | | C7 (1) | D, D1, D4, D5, D9, D10 (5) | | | | |

| ASA Ethical Guidelines (2022):<br><br>**OMB Statistical Policies**<br>**Directive 1 (2014)**<br>**Directive 2 (2008)**<br>**Directive 2a (2016)** | | A<br>Professional Integrity and Accountability | B<br>Integrity of Data and Methods | C<br>Stakeholders | D<br>Research Subjects/Data Subjects and Those Affected by Statistical Practices | E<br>Interdisciplinary Team Members | F<br>Other Practitioners/ Profession | G<br>Leader/ Supervisor/ Mentor and APPENDIX | H<br>Allegations of Potential Misconduct |
|---|---|---|---|---|---|---|---|---|---|
| SPD 2 | SECTION 1 DEVELOPMENT OF CONCEPTS, METHODS, AND DESIGN<br>Survey Planning Standard 1.1: Develop a written plan | A, A4, A11 | B, B3 | C, C1, C4 | (D) | E3 | | APPENDIX 4, 10 | |
| | Survey Design Standard 1.2: Develop a survey design | A; A2 | B2, B3 | C, C1, C4 | D2, D7 | E3 | | G5 APPENDIX 4, 10 | |
| | Survey Response Rates Standard 1.3: Design the survey to achieve the highest practical rates of response | A2 | B | C | D, D2, D6, D8, D10 | | | G5 APPENDIX 4, 10 | |
| | Pretesting Survey Systems Standard 1.4: Agencies must ensure that all components of a survey function as intended | A1, A2 | B, B1, B2, B5, B6 | C3 | D7, D8 | (E4) | | G5 | |

| **ASA Ethical Guidelines (2022):**<br><br>**OMB Statistical Policies<br>Directive 1 (2014)<br>Directive 2 (2008)<br>Directive 2a (2016)** | **A**<br>Professional Integrity and Accountability | **B**<br>Integrity of Data and Methods | **C**<br>Stakeholders | **D**<br>Research Subjects/Data Subjects and Those Affected by Statistical Practices | **E**<br>Interdisciplinary Team Members | **F**<br>Other Practitioners/ Profession | **G**<br>Leader/ Supervisor/ Mentor and APPENDIX | **H**<br>Allegations of Potential Misconduct |
|---|---|---|---|---|---|---|---|---|
| SECTION 2 COLLECTION OF DATA<br>Developing Sampling Frames<br>Standard 2.1: Ensure that the frames for the planned sample survey or census are appropriate | A; A2 | B, B1, B2, B5, B6 | C, C5 | D, D2 | | | | |
| Required Notifications to Potential Survey Respondents<br>Standard 2.2: Agencies must ensure that each collection of information instrument clearly states the reasons the information is planned to be collected; etc. | A, A2, A11 | B, B3 | C, C2, C4 | D, D7, D8, D9 | E3 | | G1 APPENDIX 10 | |
| Data Collection Methodology<br>Standard 2.3 (BALANCE QUALITY AGAINST ERROR, MINIMIZE | A, A2 | B, B1, B2, B3, B4 | C, C4 | D, D2, D6, D7, D8, D9 | E4 | | | |

| ASA Ethical Guidelines (2022): OMB Statistical Policies Directive 1 (2014) Directive 2 (2008) Directive 2a (2016) | A Professional Integrity and Accountability | B Integrity of Data and Methods | C Stakeholders | D Research Subjects/Data Subjects and Those Affected by Statistical Practices | E Interdisciplinary Team Members | F Other Practitioners/ Profession | G Leader/ Supervisor/ Mentor and APPENDIX | H Allegations of Potential Misconduct |
|---|---|---|---|---|---|---|---|---|
| BURDEN/COST) | | | | | | | | |
| SECTION 3 PROCESSING AND EDITING OF DATA Data Editing Standard 3.1: Agencies must edit data appropriately | A2, A4, A7, A11 | B, B1, B2, B5, B6 | C2, C5 | D, D8, D10 | E4 | | G5 APPENDIX 4, 6, 8, 9, 10 | H2 |
| Nonresponse Analysis and Response Rate Calculation Standard 3.2 | A2 | B, B1, B2, B6 | C4, C5 | D8, D9 | | | G5 | |
| Coding QUALITY FOR OTHERS TO USE Standard 3.3: | A2, A11 | B1, B4 | C | D8, D9, D10 | | F4 | G5 | |
| Data Protection Standard 3.4 AVOID DISCLOSURE | A3, A9, A11 | B4 | C7 | D, D1, D4, D5, D7, D9, D10, D11 | E4 | | APPENDIX 4 | H2 |

| ASA Ethical Guidelines (2022): OMB Statistical Policies Directive 1 (2014) Directive 2 (2008) Directive 2a (2016) | | A Professional Integrity and Accountability | B Integrity of Data and Methods | C Stakeholders | D Research Subjects/Data Subjects and Those Affected by Statistical Practices | E Interdisciplinary Team Members | F Other Practitioners/ Profession | G Leader/ Supervisor/ Mentor and APPENDIX | H Allegations of Potential Misconduct |
|---|---|---|---|---|---|---|---|---|---|
| | Evaluation Standard 3.5: Agencies must evaluate the quality of the data | A2, A4 | B, B1, B2, B3, B4, B5, B6 | C2, C4, C6 | D8, D10 | E3, E4 | F4 | G2 APPENDIX 10 | |
| | SECTION 4 PRODUCTION OF ESTIMATES AND PROJECTIONS Developing Estimates and Projections USING THEORY AND METHODS Standard 4.1: | A, A2 | B, B1, B2, B3, B5, B6 | C2, C4 | D8, D10 | E3, E4 | | G5 APPENDIX 4, 10 | H2 |
| | SECTION 5 DATA ANALYSIS Analysis and Report Planning Standard 5.1: Agencies must develop a plan | A2, A4 | B, B1, B2, B3 | C2, C8 | D10 | E4 | | | |
| | Inference and Comparisons Standard 5.2: USE ACCEPTABLE | A, A2, A4 | B, B1, B2, B3 | C2 | D10 | E4 | | G5 APPENDIX 4, 6, 8, 9, 10 | |

| ASA Ethical Guidelines (2022): **OMB Statistical Policies** **Directive 1 (2014)** **Directive 2 (2008)** **Directive 2a (2016)** | A Professional Integrity and Accountability | B Integrity of Data and Methods | C Stakeholders | D Research Subjects/Data Subjects and Those Affected by Statistical Practices | E Interdisciplinary Team Members | F Other Practitioners/ Profession | G Leader/ Supervisor/ Mentor and APPENDIX | H Allegations of Potential Misconduct |
|---|---|---|---|---|---|---|---|---|
| STATISTICAL PRACTICE | | | | | | | | |
| SECTION 6 REVIEW PROCEDURES  Review of Information Products Standard 6.1: REVIEW WHAT IS DISSEMINATED & COMPLY WITH OMB GLS | A, A11 | B, B5 | C | (D) | E3 | F2 | G1, G5 | |
| SECTION 7 DISSEMINATION OF INFORMATION PRODUCTS  Releasing Information Standard 7.1 (USE A PLAN THAT IS EQUITABLE) | A11 | (B4), B5, B6 | C6 | D10 | | | G5 | |

| ASA Ethical Guidelines (2022): OMB Statistical Policies Directive 1 (2014) Directive 2 (2008) Directive 2a (2016) | | A Professional Integrity and Accountability | B Integrity of Data and Methods | C Stakeholders | D Research Subjects/Data Subjects and Those Affected by Statistical Practices | E Interdisciplinary Team Members | F Other Practitioners/ Profession | G Leader/ Supervisor/ Mentor and APPENDIX | H Allegations of Potential Misconduct |
|---|---|---|---|---|---|---|---|---|---|
| | Data Protection and Disclosure Avoidance for Dissemination Standard 7.2 | A11 | B4 | C7 | D1, D4, D5, D9, D10, D11 | | | | |
| | Survey Documentation Standard 7.3 | A, A2, A3, A4 | B, B1, B2, B3, B4, B5, B6 | C, C4, C7 | D, D4, D5, D8, D11 | E3 | F4 | | |
| | Documentation and Release of Public-Use Microdata Standard 7.4: | A2 | B1, B2, G3, B4 | C4, C7 | D, D4, D5 | E3, E4 | F4 | APPENDIX 4, 10 | |
| SPD 2a | Standard A.1: Methodological Plan | A2 | B | C1 | (D1) | E4 | F4 | APPENDIX 4 | |
| | Sample Selection Standard A.2: | A2 | (B1, B2, B7) | C1, C4 | D6, D10 | | | | |
| | Interview Guide Standard A.3: | A2 | (B) | (C) | | | | | |

| ASA Ethical Guidelines (2022):<br><br>**OMB Statistical Policies**<br>**Directive 1 (2014)**<br>**Directive 2 (2008)**<br>**Directive 2a (2016)** | A<br>Professional Integrity and Accountability | B<br>Integrity of Data and Methods | C<br>Stakeholders | D<br>Research Subjects/Data Subjects and Those Affected by Statistical Practices | E<br>Interdisciplinary Team Members | F<br>Other Practitioners/ Profession | G<br>Leader/ Supervisor/ Mentor and APPENDIX | H<br>Allegations of Potential Misconduct |
|---|---|---|---|---|---|---|---|---|
| Systematic Analysis Standard A.4: | A2, A3, A4 | B1, B2, B3 | C6 | (D4) | E3 | F4 | APPENDIX 10 | |
| Transparent Analysis Standard A.5: | (A2) | B, B1, B2, B5, B6 | C, C5 | D, D2 | | | | |
| Final Reports Standard A.6: | (A2) | B1, B2, B3 | C6 | (D4) | E3 | | | |
| Reporting Results Standard A.7: | A11 | B1, B2 | C6 | D6 | E3 | F4 | APPENDIX 10 | |

NOTES. (*item*) means that this item/element must be considered while the standard is being met - e.g,. (C) suggests that responsibilities to stakeholders must be kept in mind while complying with the standard. red text highlights an element (D6, "Considers the impact of statistical practice on society, groups, and individuals. Recognizes that statistical practice could adversely affect groups or the public perception of groups, including marginalized groups. Consider approaches to minimize negative impacts in applications, or in framing results in reporting") that, if following the standard is not carefully considered, could lead to a violation of the red item. Green columns show that every standard (row) is supported by that ASA EG Principle (column).

Table 2. Alignment of ASA Ethical Guidelines (2022) with NASEM Principles and Practices (2021)

| ASA Ethical Guidelines (2022): <br><br> National Academies' Principles and Practices (2021): | A <br> Professional Integrity and Accountability | B <br> Integrity of Data and Methods | C | D <br> Research Subjects/Data Subjects and Those Affected by Statistical Practices | E <br> Interdisciplinary Team Members | F <br> Other Practitioners/ Profession | G <br> Leader/Supervisor/ Mentor and APPENDIX | H <br> Allegations of Potential Misconduct |
|---|---|---|---|---|---|---|---|---|
| **PRI1** Relevance to Policy Issues and Society | A, A2, ~A3, ~A4 | B1, B2, B3, B6 | C, C1, C2, C8 | D, D2, ~D10, ~D11 | E3, E4 | | G5 APPENDIX ~6, ~8, ~9 | |
| **PRI2** Credibility Among Data Users and Stakeholders | A, A1, A2, A4, A5, A7, A9, A10, A11, A12 | B, B1, B2, B3, B5, B6 | C, C1, C2, C3, C4, C5, C8 | D, D2, D10, D11 | E3, E4 | F5 | G1, G5 APPENDIX 1, 2, 4, 8, 9, 10, 11 | H2 |
| **PRI3** Trust Among the Public and Data Providers | A, A1, A2, A3, A4, A5, A7, A9, A10, A11, A12 | B, B1, B2, B3, B4, B5, B6 | C, C1, C2, C3, C4, C5, C7, C8 | D, D1, D2, D4, D5, D6, D7, D8, D9, D10, D11 | E3, E4 | F | G1, G5 APPENDIX 1, 2, 4, 6, 8, 9, 11 | H2 |
| **PRI4** Independence from Political and Other Undue External Influence | A, A2, A3, A4, A7 A11, A12 | B, B1, B2, B3, B5 | C2, C8 | D1, D2, D4, D5, D6, D8, D10, D11 | E4 | F4 | G1, G2, G5 APPENDIX 1, 2, 4, 6, 8, 9, 11 | H2 |
| **PRI5**: Continual Improvement and Innovation | A10 | B6 | C6 | ~D, ~D6 | | | G3 | |
| **PRA1:** A Clearly Defined and Well | (A1, A2) A4 | (B2, B3) | C1, C2 | (D5, D9) | E4 | | (G5) APPENDIX 4, 8, 9 | |

| ASA Ethical Guidelines (2022):<br><br>National Academies' Principles and Practices (2021): | A<br>Professional Integrity and Accountability | B<br>Integrity of Data and Methods | C | D<br>Research Subjects/Data Subjects and Those Affected by Statistical Practices | E<br>Interdisciplinary Team Members | F<br>Other Practitioners/ Profession | G<br>Leader/Supervisor/ Mentor and APPENDIX | H<br>Allegations of Potential Misconduct |
|---|---|---|---|---|---|---|---|---|
| Accepted Mission | | | | | | | | |
| **PRA2:** Necessary Authority and Procedures to Protect Independence | (A1, A2, A3) A4, A7, A9 | B1, B2, B3, B5 | C1, C2 | (D5, D6, D7, D8, D10, D11) | E2, E4 | (F2, F3) | G1, G2, G5 APPENDIX 1, 2, 4, 7, 8, 9, 10, 12 | H2 |
| **PRA3**: Commitment to Quality and Professional Standards of Practice | A, A1, A2, A4, A5, A7, A9, A10, A12 | B, B1, B2, B3, B5 | C, C1, C2, C8 | D, D1, D11 | E2, E3, E4 | F, F2, F3, F4, F5 | G, G1, G5 APPENDIX 1, 2, 3, 4, 6, 8, 9, 10, 12 | H, H2 |
| **PRA4:** Professional Advancement of Staff | (A, A1, A4, A6, A7, A8, A9, A11) A5, A12 | (B1) | (C2, C4) | | | (F, F2, F3) | (G1, G2) G3, G4 APPENDIX 3, 5, 7 | H, H2 |
| **PRA5:** An Active Research Program | A2 (~A1, ~A4, ~A7) | (B1, B6, B7) | (C1, C2) C6, C8 | ~D (~D5, ~D6, ~D10, ~D11) | | F4, F5 | G3, G5 APPENDIX 12, ~4, ~5, ~7, ~8, ~9 | H2 |
| **PRA6:** Strong Internal and External Evaluation Processes for an Agency's Statistical | A1 (~A4) | B, B1, B2, B3 | | D1 | (E3, E4) | F, F2, F4 | G1, G2 (G5) APPENDIX (1, 12) 10 | (H5) |

| ASA Ethical Guidelines (2022):<br><br>National Academies' Principles and Practices (2021): | A<br>Professional Integrity and Accountability | B<br>Integrity of Data and Methods | C | D<br>Research Subjects/Data Subjects and Those Affected by Statistical Practices | E<br>Interdisciplinary Team Members | F<br>Other Practitioners/ Profession | G<br>Leader/Supervisor/ Mentor and APPENDIX | H<br>Allegations of Potential Misconduct |
|---|---|---|---|---|---|---|---|---|
| Programs | | | | | | | | |
| **PRA7:** Coordination and Collaboration with Other Statistical Agencies ⇁ | A7, A11 ~A2, ~A3, ~A4 ~A5, ~A9 | ~B1, ~B2, ~B4, ~B5 | C1, ~C2, ~C8 | ~D, ~D1, ~D10, ~D11 | ~E3, ~E4 | F, F1 ~F4 | ~G, ~G1, ~G5 APPENDIX 1, 2, 4, ~6, ~8, ~9, ~10, ~11 | (~H2) |
| **PRA8:** Respect for Data Providers and Protection of Their Data | (A2) A3 | B4 (B7) | C7 | D, D1, D2, D3, D4, D5, D6, D7, D8, D9, D10, D11 | | (F4) | (G5) APPENDIX 4, 6, 8, 9 | |
| **PRA9:** Dissemination of Statistical Products That Meet Users' Needs | ~A2, ~A3, ~A4 | B1, B2, B5, B6, ~B3 | C4, C6, ~C2 | (D8), ~D6 | E3, ~E4 | | ~G1, ~G5 APPENDIX ~4, ~8, ~9, ~10, ~11 | ~H2 |
| **PRA10:** Openness About Sources and Limitations of the Data Provided | A4, A7 | B, B1, B2, B3 | C3, C4 | D8, D10 | | | APPENDIX 4, 8, 9, 10, 11 | H2 |

⇝ Note that every ASA Ethical Guideline Principle offers support for accomplishing PRA7, but the majority of this support is to inform the practitioner to take precautions so that accomplishing PRA7 does not lead the practitioner to violate any of the identified ASA Guideline elements shown.

TABLE 3: Alignment between ASA Ethical Guidelines (2022) with US Data Ethics Tenets (2020)

| ASA Ethical Guidelines (2022):<br><br>Data Ethics Tenets (2020): | A<br>Professional Integrity and Accountability | B<br>Integrity of Data and Methods | C<br>Stakeholders | D<br>Research Subjects/Data Subjects and Those Affected by Statistical Practices | E<br>Interdisciplinary Team Members | F<br>Other Practitioners/ Profession | G<br>Leader/Supervisor/ Mentor and APPENDIX | H<br>Allegations of Potential Misconduct |
|---|---|---|---|---|---|---|---|---|
| 1. Uphold applicable statutes, regulations, professional practices, and ethical standards. | A11 | (B4) | C2 (C8) | D1, D9, D11 | E; E1, E2 | F; F1 | (G; G1); G2, G5 APPENDIX 1, 2, 4, 7, 10, 11, 12 | H; H1, H2, H3, H4 |
| 2. Respect the public, individuals, and communities | A; A2, A3, A4, A5, A7 | B1, B3, B4, B5, B6 | C; C1, C2, C8 | D2, D5, D6, D7, D10, D11 | E4 | F1, F2, F3, F5 | G2, G5 APPENDIX 1, 2, 6, 8, 9 | H |
| 3. Respect privacy and confidentiality. | (A3) | B4 | C7 | D4, D5, D7, D9, D10 | | | | (H4, H5, H6 - for practitioners- not for data subjects) |
| 4. Act with honesty, integrity, and humility. | A; A1 | B1, B2, B3 | C1, C2, C3, C4, C8 | D10 | E2, E4 | F3 | G; G1, G2, G5 APPENDIX 1, 2, 4, 5, 8, 9, 10, 11 | H2 |
| 5. Hold oneself and others accountable. | A9, A12 | | C1, C2 | D; D1, D4, D10 | E3, E4 | F2, F3, F4, F5 | G; G1, G2, G5 APPENDIX 1, 2, 4, 12 | H2 |
| 6. Promote transparency. | A2, A7 | B2 | C1, C2, C4, C5, C6 | | E3 | F4, F5 | G5 APPENDIX 1, 2, 4, 8, 9, 10, 12 | (H2) |
| 7. Stay informed of developments in the fields of data management and data science. | | B6 | (C6) | | | (F4) | APPENDIX 3 (12) | |

TABLE a. Alignment between NASEM Principles and Practices and OMB Statistical Policy Directives 1, 2 and 2a

| NASEM Principles and Practices (2021):<br><br>OMB Statistical Policy:<br>Directive 1 (2014)<br>Directive 2 (2006)<br>Directive 2a (2015) | | PRI1 Relevance | PRI2 Credibility | PRI3 Trust | PRI4 Independence | PRI5 Improve/ Innovate | PRA1 Mission | PRA2 Authority | PRA3 Quality | PRA4 Staff | PRA5 Research | PRA6 Evaluation | PRA7 Collaborate | PRA8 Respect data holders | PRA9 Disseminate | PRA10 Openness |
|---|---|---|---|---|---|---|---|---|---|---|---|---|---|---|---|---|
| SPD 1 | Responsibility 1: Relevance | x | x | x | | | (~) | (~) | | | (x) | | (~) | (~) | x | (~) |
| | Responsibility 2: Credible | | x | x | x | x | x | x | x | | x | x | (~) | (x) | x, (~) | x |
| | Responsibility 3: Objective | | x | x | x | x | | | x | | x | x | (~) | (x) | x, (~) | x |
| | Responsibility 4: Trust | x | x | x | | | x | x | x | | (~) | | (~) | x | (~) | |
| SPD 2 | Standard 1.1: Develop a written plan | | x | x | | (x) | x | | x | | | | x, x | | x | |
| | Standard 1.2: develop a survey design | | x | x | | (x) | | | x | | | | x, x | | x | x |
| | Standard 1.3: design the survey to achieve the highest practical | (x) | x | x | | (x) | x | | x | | x | | x, x | (x) | x | x |

| **NASEM Principles and Practices (2021):**<br><br>**OMB Statistical Policy: Directive 1 (2014) Directive 2 (2006) Directive 2a (2015)** | PRI1 Relevance | PRI2 Credibility | PRI3 Trust | PRI4 Independence | PRI5 Improve/ Innovate | PRA1 Mission | PRA2 Authority | PRA3 Quality | PRA4 Staff | PRA5 Research | PRA6 Evaluation | PRA7 Collaborate | PRA8 Respect data holders | PRA9 Disseminate | PRA10 Openness |
|---|---|---|---|---|---|---|---|---|---|---|---|---|---|---|---|
| rates of response | | | | | | | | | | | | | | | |
| Standard 1.4: Agencies must ensure that all components of a survey function as intended | (x) | x | x | | x | x | x | x | | | x | x, x | x | x | x |
| Standard 2.1: ensure that the frames for the planned sample survey or census are appropriate | (x) | x | x | | | x | | x | | | x | x, x | | | x |
| Standard 2.2: Agencies must ensure that each collection of information instrument clearly states | | x | x | | | (x) | x | x | | | x | | x | | x |

| NASEM Principles and Practices (2021):<br><br>OMB Statistical Policy:<br>Directive 1 (2014)<br>Directive 2 (2006)<br>Directive 2a (2015) | | PRI1 Relevance | PRI2 Credibility | PRI3 Trust | PRI4 Independence | PRI5 Improve/ Innovate | PRA1 Mission | PRA2 Authority | PRA3 Quality | PRA4 Staff | PRA5 Research | PRA6 Evaluation | PRA7 Collaborate | PRA8 Respect data holders | PRA9 Disseminate | PRA10 Openness |
|---|---|---|---|---|---|---|---|---|---|---|---|---|---|---|---|---|
| | Standard 2.3 (BALANCE QUALITY AGAINST ERROR, MINIMIZE BURDEN/COST) | | x | x | (x), (x) | x | x | | x | | | x | x, x | x | | x |
| | Standard 3.1: Agencies edit data appropriately | (x) | x | x | (x), (x) | | x | (x) | x | | | x | x, x | (x) | | x |
| | Standard 3.2 Nonresponse Analysis and Response Rate Calculation | (x) | x | x | | (x) | x | | x | | | | | | (x) | x |
| | Standard 3.3: Coding QUALITY FOR OTHERS TO USE | (x) | (x) | x | | | (x), (x) | | x | | | x | x | x | x | x |
| | Standard 3.4: AVOID DISCLOSURE | | x | x | | | (x) | | x | | | x | x, x | x | x | x |

| **NASEM Principles and Practices (2021):**<br><br>**OMB Statistical Policy:**<br>**Directive 1 (2014)**<br>**Directive 2 (2006)**<br>**Directive 2a (2015)** | PRI1 Relevance | PRI2 Credibility | PRI3 Trust | PRI4 Independence | PRI5 Improve/ Innovate | PRA1 Mission | PRA2 Authority | PRA3 Quality | PRA4 Staff | PRA5 Research | PRA6 Evaluation | PRA7 Collaborate | PRA8 Respect data holders | PRA9 Disseminate | PRA10 Openness |
|---|---|---|---|---|---|---|---|---|---|---|---|---|---|---|---|
| Standard 3.5: Agencies must evaluate data quality | (x) | x | x | | | x | | x | | | x | x, x | x | x | x |
| Standard 4.1: Develop Estimates and Projections USING THEORY AND METHODS | | x | x | | (x) | | x | x | | | | x, x | | x | x |
| Standard 5.1: Agencies must develop an analysis plan | (x) | x | x | (x), (x) | (x) | | x | x | | | | x, x | | x | x |
| Standard 5.2: USE ACCEPTABLE STATISTICAL PRACTICE | | x | x | | (x), (x) | x | | x | | x, x | x | x, x | x | x | |
| Standard 6.1: REVIEW WHAT IS DISSEMINATED & COMPLY WITH OMB GLS | x | x | x | (x), (x) | x | x | (x) | x | | | x | x, x | | x | x |

| NASEM Principles and Practices (2021):<br><br>OMB Statistical Policy:<br>Directive 1 (2014)<br>Directive 2 (2006)<br>Directive 2a (2015) | | PRI1 Relevance | PRI2 Credibility | PRI3 Trust | PRI4 Independence | PRI5 Improve/ Innovate | PRA1 Mission | PRA2 Authority | PRA3 Quality | PRA4 Staff | PRA5 Research | PRA6 Evaluation | PRA7 Collaborate | PRA8 Respect data holders | PRA9 Disseminate | PRA10 Openness |
|---|---|---|---|---|---|---|---|---|---|---|---|---|---|---|---|---|
| | Standard 7.1 EQUITABLE DISSEMINATION PLAN | (x), (x) | x | x | (x) | x | | (x) | x | | | x | x, x | x | x | x |
| | Standard 7.2: Data Protection and Disclosure Avoidance for Dissemination | | x | x | | (x), (x) | | | x | | | x | x, x | x | x | x, x |
| | Standard 7.3: Survey Documentation | | x | x | (x) | x | | | x | | | x | x | | x | x |
| | Standard 7.4: Documentation and Release of Public-Use Microdata | (x) | x | x | | x | | | x | | | (x) | x, x | x, x | x | x |
| SPD 2A | Standard A.1: A methodological plan | | x | x | x | | x | | x | | | (x) | x | | x | x |
| | Standard A.2: Sample Selection | | x | x | (x) | | | | x | | | (x) | (x) | x | | x |

| **NASEM Principles and Practices (2021):**<br><br>**OMB Statistical Policy:**<br>**Directive 1 (2014)**<br>**Directive 2 (2006)**<br>**Directive 2a (2015)** | PRI1 Relevance | PRI2 Credibility | PRI3 Trust | PRI4 Independence | PRI5 Improve/ Innovate | PRA1 Mission | PRA2 Authority | PRA3 Quality | PRA4 Staff | PRA5 Research | PRA6 Evaluation | PRA7 Collaborate | PRA8 Respect data holders | PRA9 Disseminate | PRA10 Openness |
|---|---|---|---|---|---|---|---|---|---|---|---|---|---|---|---|
| Standard A.3: Interview Guide | | x | x | x | | x | | x | | | (x) | (x) | (x) | | x |
| Standard A.4: Systematic Data Analysis | | x | x | x | (x) | | (x) | x | | | | (x) | | | x |
| Standard A.5: Transparent Analysis | | x | x | x | | x | | x | | | x | x | x | x | x |
| Standard A.6: Final Reports | | x | (x) | x | | | | x | | | x | (x) | x | x | x |
| Standard A.7: Reporting Results | | x | (x) | x | | | | x | | | x | (x) | | x | x |

**NOTES:** PRA 7 is aligned with all but 2 SPD2 elements (2.1 notify/3.2 nonresponse). Two elements (3.3 code for sharing and 7.3 documentation) are simply aligned (x) whereas all other aspects of SPD2A are both aligned (x) and also have the potential for conflict (x).
**Notes:** 2 Principles and 3 practices of P&P are fully aligned with SPD 2A, but potential for conflict arises with PRA7 (coordinate/collaborate). an independent practitioner/group or project should not be required to coordinate with external agencies/groups or interests if it affects or seeks to affect the outcomes of the statistical practice covered by SPD2A.

TABLE b. Alignment of OMB Data Ethics Tenets (2020) with NASEM Principles and Practices (2021)

| National Academies' Principles and Practices (2021): \ OMB Data Ethics Tenets (2020): | T1. Uphold applicable statutes, regulations, professional practices, and ethical standards | T2. Respect the public, individuals, and communities | T3. Respect privacy and confidentiality | T4. Act with honesty, integrity, and humility | T5. Hold oneself and others accountable | T6. Promote transparency | T7. Stay informed of developments in the fields of data management and data science |
|---|---|---|---|---|---|---|---|
| **PRI1** Relevance to Policy Issues and Society | (~) | ✓, ~ | (~) | (~) | (~) | (~) | |
| **PRI2** Credibility Among Data Users and Stakeholders | ✓ | ✓ | ✓ | ✓ | ✓ | ✓ | ✓ |
| **PRI3** Trust Among the Public and Data Providers | ✓ | ✓ | ✓ | ✓ | ✓ | ✓ | ✓ |
| **PRI4** Independence from Political and Other Undue External Influence | ✓ | ✓ | ✓ | ✓ | ✓ | ✓ | |
| **PRI5** Continual Improvement and Innovation | (✓) | | ✓ | | (✓) | | ✓ |
| **PRA1:** A Clearly Defined and Well Accepted Mission | (✓) | | | | | ✓ | |
| **PRA2:** Necessary Authority and Procedures to Protect Independence | ✓ | | (✓) | ✓ | ✓ | ✓ | |
| **PRA3**: Commitment to Quality and Professional Standards of Practice | ✓ | ✓ | ✓ | ✓ | ✓ | ✓ | (✓) |
| **PRA4:** Professional Advancement of Staff | ✓ | | | ✓ | ✓ | | ✓ |

| **OMB Data Ethics Tenets (2020)**:<br><br>**National Academies' Principles and Practices (2021):** | **T1.** Uphold applicable statutes, regulations, professional practices, and ethical standards | **T2.** Respect the public, individuals, and communities | **T3.** Respect privacy and confidentiality | **T4.** Act with honesty, integrity, and humility | **T5.** Hold oneself and others accountable | **T6.** Promote transparency | **T7.** Stay informed of developments in the fields of data management and data science |
|---|---|---|---|---|---|---|---|
| **PRA5:** An Active Research Program | (✓) | | (~) | | | (✓) | ✓ |
| **PRA6:** Strong Internal and External Evaluation Processes for an Agency's Statistical Programs | ✓ | ✓ | | | ✓ | ✓ | |
| **PRA7:** Coordination and Collaboration with Other Statistical Agencies | ~ | ~ | (~) | (~) | (~) | (~) | ✓ |
| **PRA8:** Respect for Data Providers and Protection of Their Data | ✓ | ✓ | ✓ | | ✓ | ✓ | (✓) |
| **PRA9:** Dissemination of Statistical Products That Meet Users' Needs | ~ | ~ | (~) | (~) | (~) | ✓, ~ | (✓) |
| **PRA10:** Openness About Sources and Limitations of the Data Provided | ✓ | ✓ | ✓ | ✓ | ✓ | ✓ | (✓) |

TABLE c. Alignment of OMB Statistical Policy Directives 1, 2, 2a with OMB Data Ethics Tenets (2020)

| OMB Statistical Policies<br>Directive 1 (2014)<br>Directive 2 (2008)<br>Directive 2a (2016) | | **T1.** Uphold applicable statutes, regulations, professional practices, and ethical standards | **T2.** Respect the public, individuals, and communities | **T3.** Respect privacy and confidentiality | **T4.** Act with honesty, integrity, and humility | **T5.** Hold oneself and others accountable | **T6.** Promote transparency | **T7.** Stay informed of developments in the fields of data management and data science |
|---|---|---|---|---|---|---|---|---|
| SPD 1 | Responsibility 1: Produce and disseminate relevant and timely information | (x) | (x) | (x) | (x) | (x) | x | (x) |
| | Responsibility 2: Conduct credible and accurate statistical activities | x | x | x | x | x | x | (x) |
| | Responsibility 3: Conduct objective statistical activities | | | | | | x | |
| | Responsibility 4: Protect the trust of information providers by ensuring the confidentiality and exclusive statistical use of their responses | x | **x** | x | x | x | x | (x) |
| SPD 2 | SECTION 1 DEVELOPMENT OF CONCEPTS, METHODS, AND DESIGN Survey Planning Standard 1.1: Develop | x | | | x | x | x | (x) |

| OMB Statistical Policies<br>Directive 1 (2014)<br>Directive 2 (2008)<br>Directive 2a (2016) | **OMB Data Ethics Tenets (2020)**: | **T1.** Uphold applicable statutes, regulations, professional practices, and ethical standards | **T2.** Respect the public, individuals, and communities | **T3.** Respect privacy and confidentiality | **T4.** Act with honesty, integrity, and humility | **T5.** Hold oneself and others accountable | **T6.** Promote transparency | **T7.** Stay informed of developments in the fields of data management and data science |
|---|---|---|---|---|---|---|---|---|
| | a written plan | | | | | | | |
| | Survey Design Standard 1.2: Develop a survey design | x | | | x | x | | (x) |
| | Survey Response Rates Standard 1.3: Design the survey to achieve the highest practical rates of response | x | x | | x | x | x | (x) |
| | Pretesting Survey Systems Standard 1.4: Agencies must ensure that all components of a survey function as intended | x | x | (x) | x | x | x | (x) |

| OMB Statistical Policies<br>Directive 1 (2014)<br>Directive 2 (2008)<br>Directive 2a (2016) | **OMB Data Ethics Tenets (2020)**: | T1. Uphold applicable statutes, regulations, professional practices, and ethical standards | T2. Respect the public, individuals, and communities | T3. Respect privacy and confidentiality | T4. Act with honesty, integrity, and humility | T5. Hold oneself and others accountable | T6. Promote transparency | T7. Stay informed of developments in the fields of data management and data science |
|---|---|---|---|---|---|---|---|---|
| | SECTION 2 COLLECTION OF DATA<br>Developing Sampling Frames<br>Standard 2.1: Ensure that the frames for the planned sample survey or census are appropriate | x | (x) | | x | x | x | |
| | Required Notifications to Potential Survey Respondents<br>Standard 2.2: Agencies must ensure that each collection of information instrument clearly states | x | x | x | x | x | x | |
| | Data Collection Methodology<br>Standard 2.3 (BALANCE QUALITY AGAINST ERROR, MINIMIZE BURDEN/COST) | x | (x) | | x | x | | (x) |
| | SECTION 3 PROCESSING AND EDITING OF DATA<br>Data Editing Standard 3.1: Agencies | x | (x) | | x | x | x | (x) |

| OMB Statistical Policies<br>Directive 1 (2014)<br>Directive 2 (2008)<br>Directive 2a (2016) | OMB Data Ethics Tenets (2020): | T1. Uphold applicable statutes, regulations, professional practices, and ethical standards | T2. Respect the public, individuals, and communities | T3. Respect privacy and confidentiality | T4. Act with honesty, integrity, and humility | T5. Hold oneself and others accountable | T6. Promote transparency | T7. Stay informed of developments in the fields of data management and data science |
|---|---|---|---|---|---|---|---|---|
| | must edit data appropriately | | | | | | | |
| | Nonresponse Analysis and Response Rate Calculation Standard 3.2 | x | | | x | x | x | (x) |
| | Coding QUALITY FOR OTHERS TO USE Standard 3.3: | x | x | x | x | x | x | |
| | Data Protection Standard 3.4 AVOID DISCLOSURE | x | x | x | x | x | x | x |
| | Evaluation Standard 3.5: Agencies must evaluate the quality of the data | x | (x) | | x | x | x | x |

| OMB Statistical Policies<br>Directive 1 (2014)<br>Directive 2 (2008)<br>Directive 2a (2016) | | **OMB Data Ethics Tenets (2020)**: | T1. Uphold applicable statutes, regulations, professional practices, and ethical standards | T2. Respect the public, individuals, and communities | T3. Respect privacy and confidentiality | T4. Act with honesty, integrity, and humility | T5. Hold oneself and others accountable | T6. Promote transparency | T7. Stay informed of developments in the fields of data management and data science |
|---|---|---|---|---|---|---|---|---|---|
| | SECTION 4 PRODUCTION OF ESTIMATES AND PROJECTIONS Developing Estimates and Projections USING THEORY AND METHODS Standard 4.1: | | x | | | x | x | x | x |
| | SECTION 5 DATA ANALYSIS Analysis and Report Planning Standard 5.1: Agencies must develop a plan | | x | | | | x | x | |
| | Inference and Comparisons Standard 5.2: USE ACCEPTABLE STATISTICAL PRACTICE | | x | | (x) | | x | x | x |
| | SECTION 6 REVIEW PROCEDURES Review of Information Products Standard 6.1: REVIEW WHAT IS DISSEMINATED & COMPLY WITH OMB GLS | | x | (x) | x | x | x | x | |

| | OMB Data Ethics Tenets (2020): **OMB Statistical Policies** **Directive 1 (2014)** **Directive 2 (2008)** **Directive 2a (2016)** | T1. Uphold applicable statutes, regulations, professional practices, and ethical standards | T2. Respect the public, individuals, and communities | T3. Respect privacy and confidentiality | T4. Act with honesty, integrity, and humility | T5. Hold oneself and others accountable | T6. Promote transparency | T7. Stay informed of developments in the fields of data management and data science |
|---|---|---|---|---|---|---|---|---|
| | SECTION 7 DISSEMINATION OF INFORMATION PRODUCTS Releasing Information Standard 7.1 (USE A PLAN THAT IS EQUITABLE) | x | x | | x | x | x | (x) |
| | Data Protection and Disclosure Avoidance for Dissemination Standard 7.2 | x | x | x | x | x | x | |
| | Survey Documentation Standard 7.3 | x | | | | | x | |
| | Documentation and Release of Public-Use Microdata Standard 7.4: | x | x | x | | | x | |
| SPD 2a | Standard A.1: Methodological Plan | x | | | x | x | x | (x) |
| | Sample Selection Standard A.2: | x | (x) | | x | x | x | |

| **OMB Statistical Policies**<br>**Directive 1 (2014)**<br>**Directive 2 (2008)**<br>**Directive 2a (2016)** | **OMB Data Ethics Tenets (2020):** | **T1.** Uphold applicable statutes, regulations, professional practices, and ethical standards | **T2.** Respect the public, individuals, and communities | **T3.** Respect privacy and confidentiality | **T4.** Act with honesty, integrity, and humility | **T5.** Hold oneself and others accountable | **T6.** Promote transparency | **T7.** Stay informed of developments in the fields of data management and data science |
|---|---|---|---|---|---|---|---|---|
| | Interview Guide Standard A.3: | (x) | | | | x | x | |
| | Systematic Analysis Standard A.4: | x | | | | | | (x) |
| | Transparent Analysis Standard A.5: | x | x | x | x | x | x | |
| | Final Reports Standard A.6: | x | | | | x | x | |
| | Reporting Results Standard A.7: | x | (x) | (x) | | x | x | |

NOTE: DET 1 is aligned with all of the SPD 1, 2, 2A elements. R2, R4, standard 1.4 and standard 3.4 are aligned with all DET tenets. Cautions (potentials for conflict) arise relating to dissemination responsibilities (R1) and standards (6.1, A7) because DET T1, T2, and T3 charge the statistical practitioner with the obligation to respect the public, individuals, and communities, as well as privacy and confidentiality. T1 charges the practitioner with following ethical standards, and the ASA Ethical Guidelines for Statistical Practice include multiple instances of obligations to consider the potential for negative impacts of reports on individuals, groups, and communities.

TABLE d. Alignment of OMB Statistical Policy Directive 1 (2014) with OMB Data Ethics Tenets (2020)

| OMB Statistical Policy Directive 1 2014: <br><br> OMB OMB Data Ethics Tenets 2020: | R1 Produce and disseminate relevant and timely information. | R2 Conduct credible and accurate statistical activities | R3 Conduct objective statistical activities | R4 Protect the trust of information providers by ensuring the confidentiality and exclusive statistical use of their responses |
|---|---|---|---|---|
| **T1.** Uphold applicable statutes, regulations, professional practices, and ethical standards. | (~) | ✓ | ✓ | ✓ |
| **T2.** Respect the public, individuals, and communities | (~) | ✓ | ✓ | ✓ |
| **T3.** Respect privacy and confidentiality. | (~) | ✓ | ✓ | ✓ |
| **T4.** Act with honesty, integrity, and humility. | (~) | ✓ | ✓ | ✓ |
| **T5.** Hold oneself and others accountable. | (~) | ✓ | ✓ | ✓ |
| **T6.** Promote transparency. | ✓ | ✓ | ✓ | ✓ |
| **T7.** Stay informed of developments in the fields of data management and data science. | (✓) | (✓) |  | (✓) |

TABLE e. SPD3 Statistical Policy Directive No. 3: Compilation, Release, and Evaluation of Principal Federal Economic Indicators (1985) WITH ASA EGS (2022)

| ASA Ethical Guidelines (2022):<br><br>SPD3 (1985): | A<br>Professional Integrity and Accountability | B<br>Integrity of Data and Methods | C<br>Stakeholders | D<br>Research Subjects/Data Subjects and Those Affected by Statistical Practices | E<br>Interdisciplinary Team Members | F<br>Other Practitioners/ Profession | G<br>Leader/ Supervisor/ Mentor and APPENDIX | H<br>Allegations of Potential Misconduct |
|---|---|---|---|---|---|---|---|---|
| 1 Designation of Principal Indicators | A1 | B1, B2, B3, B6 | C1, C4, C5 | D6, D10, D11 | E4 | F4 | APPENDIX 4 | |
| 2 Prompt Release | A2 , A11 | B, B2 | (C4) | D8, D10 | | | | |
| 3 Release Schedule | A11 | | (C) | | E3 | | | |
| 4 Announcement of Changes | A11 | B5, B6 | C2, C5 | | E3 | | | |
| 5 Release Procedure | A11 | (B4) | C6 | D10 | | | | |
| 6 Preliminary Estimates and Revisions | A, A11 | B, B2, B3, B4, B5, B6 | C2 | D10 | E3 | F4 | G5 | |
| 7 Granting of Exceptions | (A9), A11 | B4 | C2 | | E3, E4 | | G5 | H2 |
| 8 Performance Evaluation | A11 | B, B1, B2, B3, B6 | (C) | | | | G5 | |

NOTES. (*item*) means that this item/element must be considered while the standard is being met - e.g,. (C) suggests that responsibilities to stakeholders must be kept in mind while complying with the standard. Green columns show that every standard (row) is supported by that ASA EG Principle (column).

TABLE f. SPD4 Statistical Policy Directive No. 4: Release and Dissemination of Statistical Products Produced by Federal Statistical Agencies (2008) WITH ASA EGS (2022)

| ASA Ethical Guidelines (2022):<br><br>SPD4 (2008): | PREAMBLE | A Professional Integrity and Accountability | B Integrity of Data and Methods | C Stakeholders | D Research Subjects/Data Subjects and Those Affected by Statistical Practices | E Interdisciplinary Team Members | F Other Practitioners/ Profession | G Leader/ Supervisor/ Mentor and APPENDIX | H Allegations of Potential Misconduct |
|---|---|---|---|---|---|---|---|---|---|
| 1. implement consistent with relevant Statistical Policy Directives | | A11 | | | | | | | |
| 2. Definition of statistical products | x | | | | | | | | |
| 3. Definition of statistical purpose | x | | | | | | | | |
| 4. Timely release | | A2, A11 | B, B1, B2 | (C), (C4), C5, C6, C7 | D8, D10 | | | | |
| 5. Timely public notification | | | | | (D10) | | | | |

| | | | | | | | | |
|---|---|---|---|---|---|---|---|---|
| 6. Dissemination | A11 | B4, B5 | C6 | D10, D11 | E3 | F4 | G5 APPENDIX 4, 10 | H2 |
| 7. Announcement of Changes | (A2, A11) | | (C, C4) | | | | | |
| 8. Revisions and Corrections of Data | | B2, B4, B5, B6 | C2 | D10 | E3 | | | |
| 9. Exceptions | (A9), A11 | B4 | C2 | | E3, E4 | | G5 | H2 |

NOTES. (*item*) means that this item/element must be considered while the standard is being met - e.g,. (C) suggests that responsibilities to stakeholders must be kept in mind while complying with the standard. Green columns show that every standard (row) is supported by that ASA EG Principle (column). NB, the Preamble is ignored for this assessment.

TABLE g. QUESTIONS AND ANSWERS WHEN DESIGNING SURVEYS FOR INFORMATION COLLECTIONS (2006/2016) x ASA EG (2022)

| ASA Ethical Guidelines (2022): <br><br> **Survey Design (2006/2016)):** | A<br>Professional Integrity and Accountability | B<br>Integrity of Data and Methods | C<br>Stakeholders | D<br>Research Subjects/Data Subjects and Those Affected by Statistical Practices | E<br>Interdisciplinary Team Members | F<br>Other Practitioners/ Profession | G<br>Leader/Supervisor/ Mentor and APPENDIX | H<br>Allegations of Potential Misconduct |
|---|---|---|---|---|---|---|---|---|
| OVERALL DOCUMENT: | A2, A11 | B2 |  | D11 |  |  | APPENDIX 4 |  |
| The reasons the information is to be collected[30] | (A11) | B3 | C1 | (D10) |  |  |  |  |
| The way the information will be used to further agency purposes and serve agency needs[31] | (A4) A11 | B3 | C1, C4 |  |  |  |  |  |
| An estimate of the average burden of the collection and whom to contact about the estimate[32] | (A8), (A11) |  |  |  |  |  |  |  |

---

[30] (44 U.S.C. § 3506(c)(1)(B)(iii)(I); 5 C.F.R. § 1320.8(b)(3)(i))

[31] (44 U.S.C. § 3506(c)(1)(B)(iii)(II); 5 C.F.R. § 1320.8(b)(3)(ii))

[32] (44 U.S.C. § 3506(c)(1)(B)(iii)(III); 5 C.F.R. § 1320.8(b)(3)(iii))

| ASA Ethical Guidelines (2022):<br><br>**Survey Design (2006/2016)):** | **A**<br>Professional Integrity and Accountability | **B**<br>Integrity of Data and Methods | **C**<br>Stakeholders | **D**<br>Research Subjects/Data Subjects and Those Affected by Statistical Practices | **E**<br>Interdisciplinary Team Members | **F**<br>Other Practitioners/ Profession | **G**<br>Leader/Supervisor/ Mentor and APPENDIX | **H**<br>Allegations of Potential Misconduct |
|---|---|---|---|---|---|---|---|---|
| Whether responses to the collection of information are voluntary or mandatory, or required to obtain a benefit[33] | A3, A11 | B3 | C2 | | | | | |
| The nature and extent of confidentiality to be provided, if any.[34] (This provision was included in the regulation as a necessary component of telling the respondent of "the way such information is to be used".)[35] | (A11) | B4 | C7 | D1, D4, D9 | | | | |

---

[33] (44 U.S.C. § 3506(c)(1)(B)(iii)(IV); 5 C.F.R. § 1320.8(b)(3)(iv))

[34] (5 C.F.R. § 1320.8(b)(3)(v)

[35] (44 U.S.C. § 3506(c)(1)(B)(iii)(II); see 5 C.F.R. § 1320.8(b)(3)(ii)))

| ASA Ethical Guidelines (2022):<br><br>**Survey Design (2006/2016)):** | A<br>Professional Integrity and Accountability | B<br>Integrity of Data and Methods | C<br>Stakeholders | D<br>Research Subjects/Data Subjects and Those Affected by Statistical Practices | E<br>Interdisciplinary Team Members | F<br>Other Practitioners/ Profession | G<br>Leader/Supervisor/ Mentor and APPENDIX | H<br>Allegations of Potential Misconduct |
|---|---|---|---|---|---|---|---|---|
| The duration of respondents' expected involvement (e.g., if this is a longitudinal survey, they should be informed that they will be contacted in the future) | (A8) | | | | | | | |
| If the agency is collecting "sensitive information," respondents should be informed about what type(s) of sensitive information will be requested. | (A11) | | | D4, D5, D8 | | | | |
| NOTES. (*item*) means that this item/element must be considered, or is implied, while the standard is being met - e.g,. (A11) suggests that complying with the standard is implied by A11, ("Follows applicable policies..."). "(A8)" means that following the standard will result in dignity and fair treatment of all people (A8). Green columns show that every standard (row) is supported by that ASA EG Principle (column). ||||||||